\documentclass[12pt,preprint]{aastex}
\usepackage{epsfig}
\usepackage{longtable}
\def\gsim{\lower 2pt \hbox{$\, \buildrel {\scriptstyle >}\over
{\scriptstyle \sim}\,$}}
\def\lsim{\lower 2pt \hbox{$\, \buildrel {\scriptstyle <}\over
{\scriptstyle \sim}\,$}}

\def\chandra{{\sl Chandra}}

\newcommand{\as}{$^{\prime\prime}~$}

\newcommand{\etal}{et al.~}

\def\xs{IC~10}

\newcommand{\xmm}{{\em XMM-Newton}}

\shortauthors{}
\shorttitle{XMM-Newton Observation of \xs}

\begin{document}
\slugcomment{Draft version: \today}
\title{An {\it XMM-Newton} and {\it Chandra} Study of the Starburst Galaxy IC 10}

\author{Q. Daniel Wang, Katherine E. Whitaker, \& Rosa Williams\footnote{Currently at Astronomy Department, University of Illinois at Urbana-Champaign, 1002 West Green Street, Urbana, IL 61801 }}

\affil{Astronomy Department, University of Massachusetts, Amherst, MA 01003,USA}
\begin{abstract}

We present an X-ray study of our nearest
starburst galaxy IC 10, based on {\em XMM-Newton} and \chandra\ observations.  
A list of 73 \xmm\ and 28 \chandra\ detections of point-like X-ray sources in the
field is provided; a substantial fraction of them are 
likely stellar objects in the Milky Way due to the low Galactic latitude
location of IC~10. The brightest source 
in the IC~10 field, X-1, has a mean 0.3-8.0 keV luminosity of 
$\sim 1.2 \times 10^{38} {\rm~erg~s^{-1}}$
and shows a large variation 
by a factor of up to $\sim 6$ on time scales 
of $\sim 10^4$ s  during the \xmm\ observation. The X-ray spectra 
of the source indicate
the presence of a multi-color blackbody accretion disk with an inner disk 
temperature $T_{in} \approx 1.1$ keV. These
results are consistent with the interpretation of the source as
a stellar mass black hole, probably accreting from a Wolf-Rayet star companion.
We infer the mass of this black hole to be $\sim 4$ 
M$_\odot$ if it is not spinning, or a factor of up to $\sim 6$ 
higher if there is significant spinning. We also detect an apparent 
diffuse soft X-ray emission 
component of IC~10. An effective method is devised 
to remove the X-ray CCD-readout streaks of X-1 that
strongly affect the study of the diffuse component in the \xmm\ and 
\chandra\ observations. We find that the diffuse X-ray morphology is oriented
along the optical body of the galaxy and is chiefly associated with
starburst regions. The diffuse component can be characterized by an optically 
thin thermal plasma with a mean temperature of $\sim 4 \times 10^6$ K and 
a 0.5-2 keV luminosity of $\sim 8 \times 10^{37} {\rm~erg~s^{-1}}$, 
representing
only a small fraction of the expected mechanical energy
inputs from massive stars in the galaxy. There is evidence that the
hot gas is driving outflows from the starburst regions; therefore, 
the bulk of the energy inputs may be released in a galactic wind.

\end{abstract}

\section{Introduction}

X-ray observations are the most sensitive tool available for the study of  
massive star end products, both stellar and interstellar. 
IC~10 is our nearest laboratory for such a study in 
a dwarf starburst galaxy. 
This galaxy is particularly known for its
large number of Wolf-Rayet (W-R) stars; the global surface density,
$\sim 40$ W-R kpc$^{-2}$ over an area of $\sim 2.5 {\rm~kpc^{-2}}$, is about 20 times 
higher than in the Large Magellanic Cloud (Richer et al. 2001; Massey \& Holmes 2002).
This large population of W-R stars is very unusual for the low metallicity of \xs\
($\sim 15\%$ solar; e.g., Lequeux et al. 1979; Crowther et al. 2003).
The galaxy also contains a number of energetic interstellar shells;
their origins are, however, unclear (e.g., Wilcots \& Miller 1998; 
Yang \& Skillman 1993). The presence of an enigmatic radio emission 
shell of diameter $\sim 140$~pc around the brightest X-ray source \xs\ X-1
is particularly interesting
\citep{Yang93, Bauer04}. 
The radio emission is predominantly nonthermal, as indicated by its
steep spectrum. The required presence of
an extremely powerful source of relativistic particles is {\sl not}
expected for a typical shell-like supernova remnant  on such a large scale 
or for a young superbubble 
produced by an OB association, which would generate overwhelming amounts
of ionizing radiation. Indeed, the shell shows only weak H$\alpha$ (or
thermal) emission \citep{Yang93}. It  has been suggested that 
the nonthermal radio shell may be
powered by \xs\ X-1, similar to the link between the Galactic X-ray binary SS~433 
and its surrounding nonthermal structure W50 (diameter $\sim 50 \times 100$ pc), albeit 
on greater physical and energetic scales \citep{Brandt97, Bauer04}.

\xs\ X-1 is the focus of the previous X-ray studies by \citet{Brandt97}
and \citet{Bauer04}.
They  find that the source is spatially coincident with the
W-R star [MAC92] 17A in \xs\ (Massey, Armanroff, \& Conti 1992) 
and speculate that the accreting
compact object is a black hole (BH) based on its high X-ray 
luminosity, strong variability, and the high mass of this 
likely companion star.
It is of great interest to confirm the identity of this compact object,
which is comparable in many ways to Cyg X-3 --- a candidate for a 
rare class of compact star/W-R binaries, which may shed light onto the differing pre-supernova evolutionary 
scenarios for very massive stars (Clark \& Crowther 2004). 

Here we present our \xmm\ observations of \xs, together with a re-analysis of
the {\sl Chandra} data studied  by \citet{Bauer04}. These two complementary data sets together give us 
a comprehensive X-ray view of the galaxy.
The \chandra\ data provide us with arcsecond spatial resolution,
important for isolating point-like sources, especially in regions
close to X-1. But the data, taken with the detector ACIS-S in 
a (500 pixel) sub-array mode, covers only a 
portion of the galaxy.  In addition, the significant pile-up of X-1 adds a degree
of spectral complexity\citep{Davis01, Bauer04}. Furthermore,
the CCD-readout streak of the source significantly contaminates the 
``diffuse'' X-ray component observed.
The \xmm\ observations provide both a higher effective photon-collecting area 
and a larger 
field coverage than the \chandra\ observation, 
albeit with a substantially poorer resolution (FWHM
$\sim 13$\as). While the 
pile-up of X-1 is not an issue for the \xmm\ observations, 
the CCD-readout streak contamination
is serious. Therefore, we need to remove
these streaks in order to correctly map out the diffuse X-ray component
of \xs.

In this work, we adopt the distance of \xs\ to be $\sim 0.7$ Mpc
(hence $1^\prime$ = 204 pc; 
Demers, Battinelli, \& Letarte  2004 and references therein).
While data points in plots are shown  all with $1\sigma$ error bars, 
our quoted parameter uncertainty ranges are at the 90\% confidence level
unless otherwise noted.

\section{Observations and Data Calibration}

\subsection{\xmm\ Data}

The \xmm\ observations were taken on July 3, 2003 with a total
exposure time of $\sim$ 45 ks for the EPIC-MOS cameras and $\sim$42 ks
for the EPIC-PN camera. We use only the data
from the PN camera (with a thin optical blocking filter) because of its high
sensitivity to soft X-rays in the study of diffuse emission
and include the data from the MOS cameras in the analysis of X-1.
The data from the RGS and OM provide no useful information here and are 
thus not presented here.

We calibrate the data with the SAS software (version 6.1.0), together with
the latest calibration files. For the MOS data,
we include only events with patterns 0 through 12, and the 
standard MOS flags (XMMEA\_EM).  The PN events are filtered to 
include only events with patterns 0 through 4, and the 
PN flags (XMMEA\_EP, XMMEA\_2, and XMMEA\_3).
We exclude time intervals with strong flares to reduce 
the non-cosmic X-ray background, which is due to cosmic-ray 
induced events, instrumental fluorescence, and soft protons that are funneled 
towards the detectors by the X-ray mirrors. We find that the flaring is 
consistent in the 0.3-2 keV and 10-15 keV bands, in which the light curves
of the observations are examined. We thus remove time intervals with 
significant count rate (CR) deviations from the mean in the 0.3-2 keV band.
The chosen threshold is 0.8 counts~s$^{-1}$ for the MOS and 2.0 counts~s$^{-1}$ for 
the PN data. This filtering leaves a total accepted 
net exposure time of $\sim$ 35.5 ks for each of the two MOS cameras and 
$\sim$ 32 ks for the PN camera.  

\begin{figure}[!hbt]
  \centerline{
      \epsfig{figure=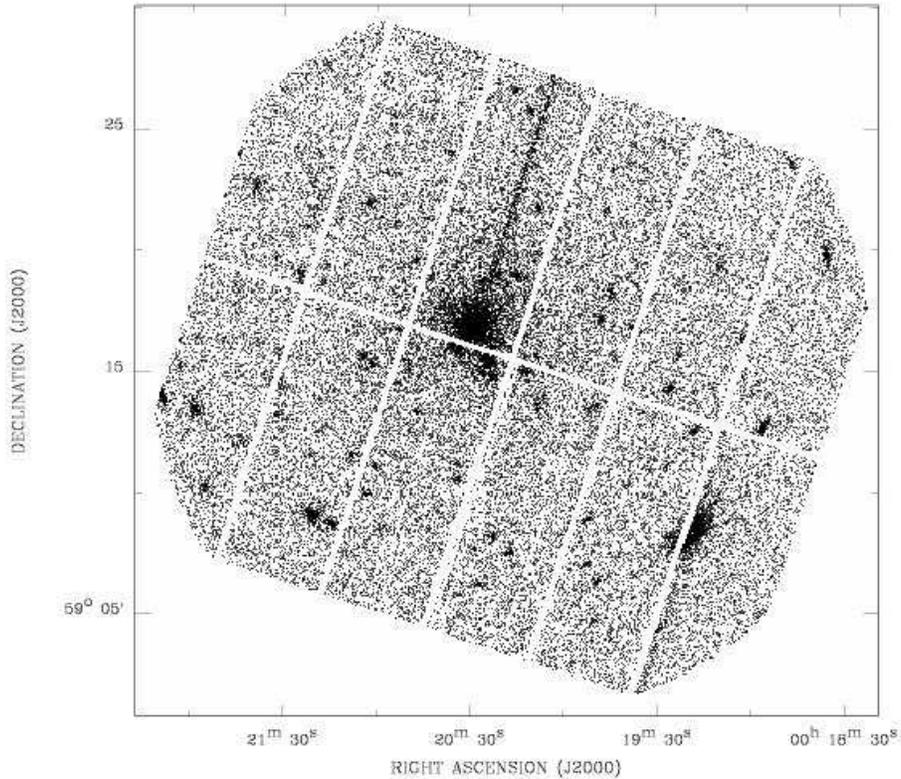,width=0.65\textwidth,angle=90}
    }
  \caption{\xmm\ EPIC-PN image in the 0.5-7.5 keV band. 
The image's northwest and southeast edges (each 1\farcm5 wide) 
have been removed for high instrument background and are not used in our 
data analysis. The CCD-readout streak of the brightest source
X-1 is apparent in chip \# 4.  The bright source in the lower right chip
also produces a faint but visible streak, which does not significantly 
affect our analysis and is thus not removed here.
    \label{fig:im_pn_raw1-4}}
\end{figure}

For the PN data, we construct maps in the 0.5-1, 1-2, 2-4.5, and 4.5-7.5 keV 
bands. We have also examined the data in the 0.2-0.5 keV band, but find
that it contains too many artifacts and little X-ray emission from
\xs. We generate corresponding effective exposure images, 
using the SAS program $eexpmap$.
These exposure maps correct for various instrumental artifacts, including
bad pixels and columns, the detector quantum efficiency non-uniformity, 
filter transmission, and telescope vignetting. Such a correction
is made on a count image after the removal of a background,
which is partly due to non-cosmic X-rays not vignetted by the
telescope. We generate the background data, using the {\sl skycast} program 
and the ``blank sky'' database --- source-removed
high Galactic latitude observations with a total exposure of 
352 ks  \citep{Read03}.  The background data are processed with 
the same filters as used for the \xs\  observations. 

\subsection{CCD-readout  Streak Removal}

\begin{figure}[!hbt]
  \centerline{
      \epsfig{figure=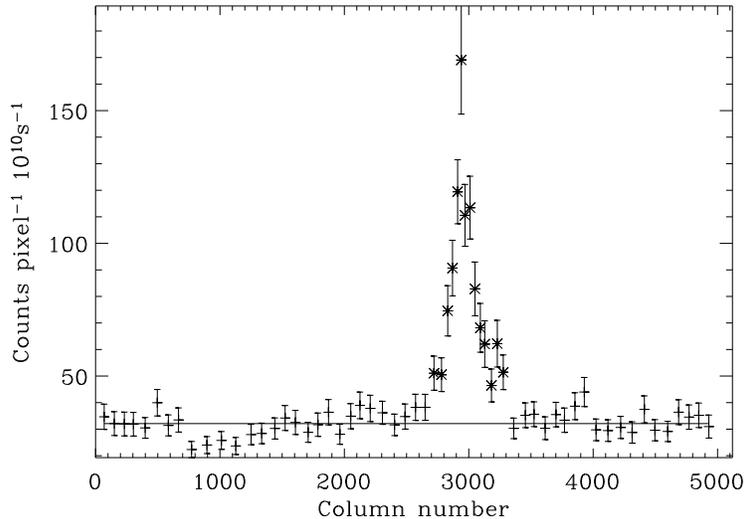,width=0.65\textwidth}
    }
  \caption{\xmm\ EPIC-PN intensity profile across the CCD chip \# 4 along
the direction perpendicular to the CCD readout columns 
(Fig.~\ref{fig:im_pn_raw1-4}). The regions around X-1 and other 
sources are removed in the intensity calculation. The columns marked with
``$\times$'' are considered to be significantly affected by the streak.
The intensity excess of the columns above the average (the horizontal line) 
over the unaffected ranges is considered to be the streak contribution.
 \label{fig:pn_st}}
\end{figure}

\begin{figure}[!htb]
  \centerline{
      \epsfig{figure=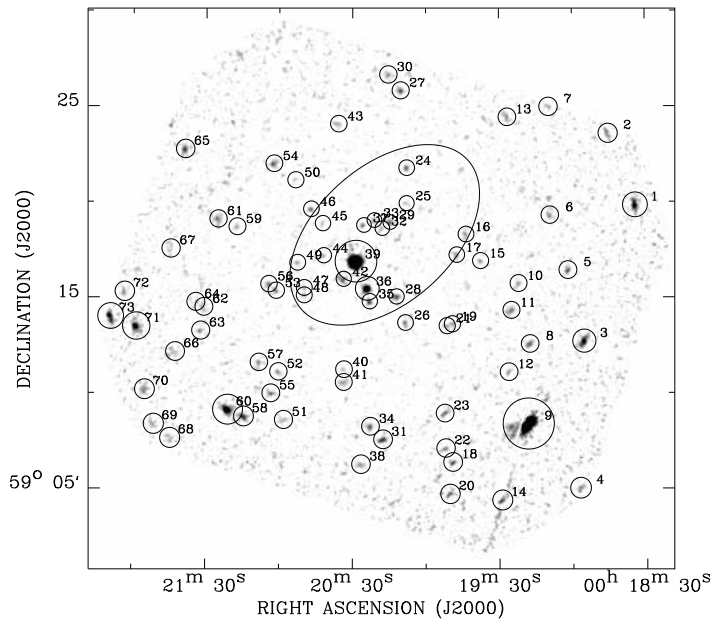,width=0.55\textwidth,angle=90}
    }
  \caption{EPIC-PN intensity image in the 0.5-7.5 keV band after
the CCD streak removal (Fig.~\ref{fig:im_pn_raw1-4}) and flat-fielding. 
The circles outline the source
regions (two times the 50\% EER), which are to be removed for diffuse X-ray analysis. An adaptively
smoothed background has also been subtracted from
the image to highlight discrete sources. The source 
numbers (Table 1) are  marked. The ellipse outlines the  2MASS ``total" size of \xs\
in the K$_s$ band (Jarrett et al. 2003).
    \label{fig:sou_pn}}
\end{figure}

The most outstanding artifact in the \xmm\ data 
is the southeast-northwest streak 
(Fig.~\ref{fig:im_pn_raw1-4}), which was caused  by 
the continuous exposure of the PN CCD (\# 4) to X-1 during its
readout. This streak spreads the counts 
(statistically) evenly into columns along
the read-out direction of the CCD. Such a streak is typically
removed by replacing the affected columns with events randomly generated from 
an estimated background spectrum. This simple cosmetic
repair, however,
would also remove any sources and/or extended structures in the columns and 
would create new artifacts, especially in regions near X-1. We have
developed a more effective streak removal method. We calculate the
average intensity distribution as a function of the column number across
the CCD chip (Fig.~\ref{fig:pn_st}), 
after excluding regions that are significantly affected by diffuse or 
scattered X-1 emission as well as individual detected sources 
(to be discussed later). 
The calculated streak contribution is then cast into the sky coordinates. 
The resultant streak image can then be subtracted from the observed 
X-ray intensity image. Fig.~\ref{fig:sou_pn} 
shows such a ``streak-free'', background-removed, and 
flat-fielded intensity image in the
0.5-4.5 keV band.  The streak produces no apparent effect in lower or higher
energy bands. A streak image in a sub-energy band can easily be produced
with a normalization of the relative intensity
of the streak.

\begin{figure}[!htb]
  \centerline{
      \epsfig{figure=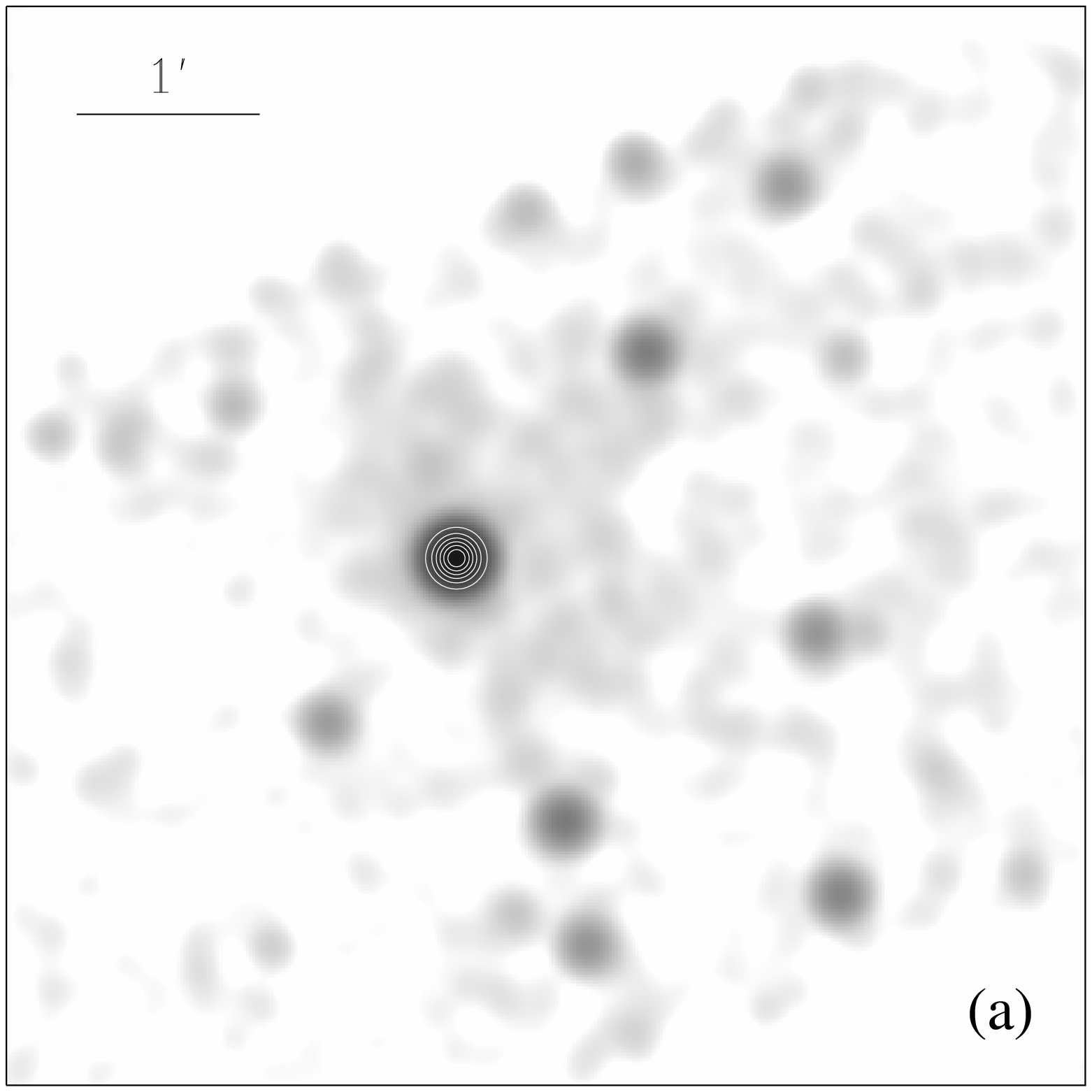,width=0.45\textwidth,angle=0}
      \epsfig{figure=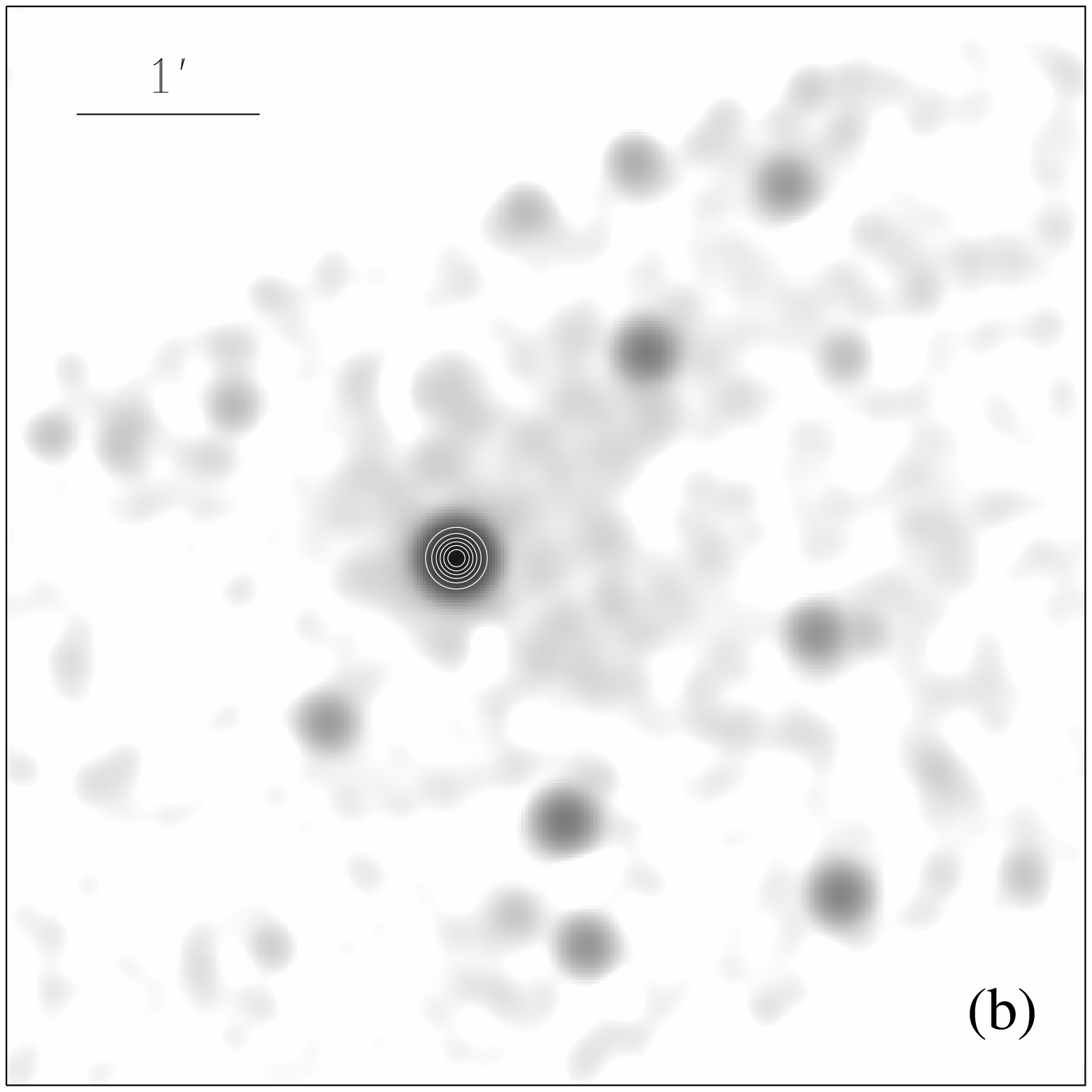,width=0.45\textwidth,angle=0}
    }
  \caption{\chandra\ ACIS-S images in the 0.7-3 keV band
(a) before and (b) after the CCD-readout streak removal.
The images have been smoothed with a Gaussian of FWHM=12\as.
    \label{fig:acis_st}}
\end{figure}

\subsection{\chandra\ Data}

The \chandra\ observation, taken on March 12, 2003 with an exposure of 29 ks, 
has already been described by \citet{Bauer04}. We reprocess
the data, using the latest CIAO software (version 3.2.1) and calibration database
(version 3.0.0), following the same procedure as detailed in Wang et al.
(2003). We create ACIS-S images and corresponding exposure (flat-fielding) images
in the 0.3-0.7, 0.7-1.5, 1.5-3, and 3-7 keV bands.

A CCD-readout streak similar to that seen in the PN data is also found in the \chandra\ ACIS-S image 
 in the 0.7-3 keV range (Fig.~\ref{fig:acis_st}a). We remove 
this  CCD-readout streak in the same way as used for the PN data. 
The total number of counts contributed by the streak is only about 92, a 
$5\sigma$ excess 
above the local background. The streak-removed image is 
shown in Fig.~\ref{fig:acis_st}b.

The \chandra\ observation has a superb on-axis resolution of 90\% EER$=1-2$\as within $\sim 4^\prime$.
However, the PSF degrades steeply at larger off-axis angles. The \xmm\ PSF
is relatively  uniform across much of the PN field of view. 
Therefore, our use of the \chandra\  data is limited within 
the central region as shown in Fig.~\ref{fig:sou_acis}. 

\section{Data Analysis and Results}

\subsection{Discrete Sources}

We use an IDL-based program to detect sources in both the \chandra\
ACIS-S and the \xmm\ PN data. As in the previous applications 
(e.g., Wang et al. 2003; Wang 2004a), this program, optimized to
detect point-like sources, uses a combination of
detection algorithms: 
wavelet, sliding-box, and maximum likelihood centroid fitting.
First, we construct the `Mexican cap'  wavelet images on
scales of 1, 2, 4 and 8 image bins (bin size $=4$\as for the PN and 0.5\as for
the ACIS-S). 
On each scale we search for source candidates corresponding to 
local maxima with signal-to-noise ratios $S/N > 2.5$. 
Next, we apply a map detection (`sliding box' method) with a background 
map produced by excising the source candidates and adaptively smoothing 
the map to achieve a local count-to-noise ratio 
greater than 10. Finally, the sources detected with the map 
method are analyzed by a maximum likelihood algorithm, using both the 
background map and an approximate Gaussian PSF.  This analysis gives
optimal source positions and their $1\sigma$ errors.
The source detection in both the map detection and the maximum 
likelihood analysis is based on data within the 50\% PSF
energy-encircled radius (EER) for the PN data and the 90\% EER for the 
ACIS-S data. The final accepted sources all have the individual
false detection probability $P\le 10^{-6}$ (ACIS-S) and $10^{-7}$ (PN).  
The processed PN data still contain various residual artifacts
(more significant than in the ACIS-S data);
the chosen stringent detection threshold minimizes their effect in the
source detection.

The source detections are carried out in 
the broad (B), soft (S), and hard (H) bands, defined differently for
the ACIS-S and PN data sets in the notes
to Tables 1 and 2. For
each data set, the detected sources in the three bands are 
merged together. Multiple detections with overlapping 2$\sigma$ centroid 
error circles are considered to be the same source,
and the centroid with the smallest
error is adopted.

\begin{figure}[!htb]
  \centerline{
      \epsfig{figure=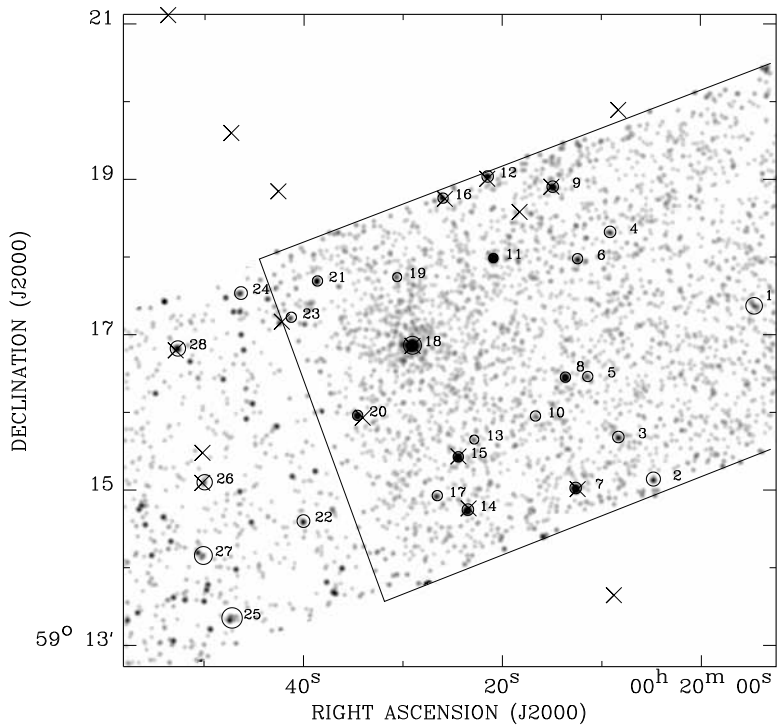,width=0.7\textwidth,angle=0}
    }
  \caption{\chandra\ ACIS-S intensity image in the 0.3-7 keV band after 
the CCD streak is removed and flat-fielded. The field covered by the on-axis
S7 chip is outlined; the remaining (low left) coverage of the image is 
from part of the S2 chip. The circles represent the source
regions  (two times the 90\% EER), which are to be removed for diffuse X-ray analysis. The 
source numbers (Table 2) are also marked. Positions of \xmm\ PN sources 
in the field are represented by {\sl crosses}.
    \label{fig:sou_acis}}
\end{figure}

Tables 1 and 2 summarize the results from our 
source detection in the two data sets. 
For ease of reference, we will refer to X-ray sources detected in the \xmm\ 
PN and \chandra\ ACIS data sets with the prefixes XP and XA,
respectively (e.g., XP-13). 
The source locations are marked in the X-ray intensity 
images (Figs.~\ref{fig:sou_pn} and \ref{fig:sou_acis}). 
The detection reveals many more
sources in the \chandra\ data than in the \xmm\ data in the overlapping
region around X-1. Part of this
difference is due to the limited spatial resolution of the \xmm\ instrument;
the source detection sensitivity is very much affected by the PSF of X-1 
and the presence of the diffuse emission, as well as source confusion. 
Nevertheless, most bright X-ray sources
in the \chandra\ field are also detected with the \xmm\ data. Several
relatively bright sources, such as XA-13, XA-17, and XP-49 (the PN source
next to XA-35 in Fig.~\ref{fig:sou_acis}), should have been detected in 
both data sets. The fact that they appear only in one data set indicates
their strong variability from one observation to another. 

\begin{figure}[!htb]
  \centerline{
      \epsfig{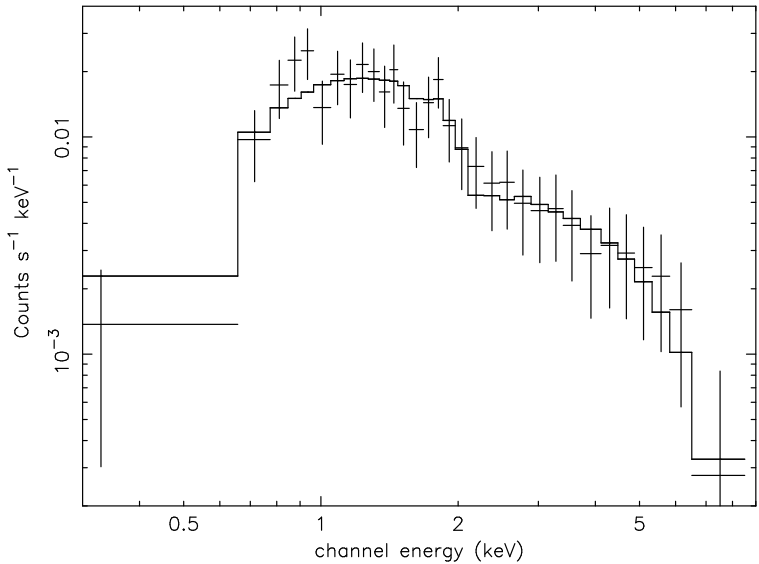}
    }
  \caption{Accumulated ACIS-S spectrum of \chandra-detected sources
on the S3 chip (Fig.~\ref{fig:sou_acis}), together with the best-fit power
law model. The spectrum is binned with a minimum number of counts to be 25.
    \label{fig:spec_sou_acis}}
\end{figure}

We may further use the \xmm\ and \chandra\ data sets to provide 
limited X-ray spectral information on the
detected sources. Most of the \xmm\ detected sources are 
apparently outside the optical/near-IR 
main body of \xs\ and are probably unrelated to the galaxy
(Fig.~\ref{fig:sou_pn}). The \chandra\
detected sources, on the other hand, are more likely to be associated
with the galaxy. We thus extract an accumulated ACIS-S spectrum of the
sources located in the S3 chip, excluding sources that are located at 
the edges (i.e., XA-12, 16, and 23; Fig.~\ref{fig:sou_acis}). This spectrum
can be characterized ($\chi^2/dof = 11/29$) 
by a power law with a photon index of
1.2(0.89 - 1.6) and an absorption column density 
$N_H = 1.8(0.9-2.9)\times 10^{21} {\rm~cm^{-2}}$ 
(Fig.~\ref{fig:spec_sou_acis}).
Although the constraints on the spectral parameters are generally
weak, the index is consistent with the value ($\sim 1-2$) 
expected for typical high-mass X-ray binaries. The $N_H$ is 
considerably lower than that expected total absorption toward sources
inside \xs; the Galactic \ion{H}{1} absorption alone is $N_{HI} 
= 4.8 \times 10^{21} {\rm~cm^{-2}}$ (Stark et al. 1992). The absorption-corrected
luminosity of the sources is $3 \times 10^{37} {\rm~erg~s^{-1}}$ in the 0.3-8 keV band.
Therefore, the
spectrum, especially in the low energy part, 
may still significantly be contaminated by Galactic objects. 

For relatively bright sources, we give hardness ratios to 
constrain individual 
spectral properties (Tables 1 and 2). Most of the calculated ratios 
are consistent with a power law that has an index in the range of 
1-3, together with a foreground absorption $N_H = (0.3 - 3) \times 10^{22}
{\rm~cm^{-2}}$. Such sources are probably X-ray binaries in 
\xs\ or background AGNs. Soft X-ray sources, especially
those with HR1 values $\lesssim 0.3$ (Table 1), should 
be in our Galaxy (\S 4.1),
because the inferred $N_H$ is less than the accepted Galactic absorption.
Most of the other sources are too faint to be constrained individually. 
A useful, though crude, conversion factor from a count rate (in either 
Table 1 or 2) to an absorption-corrected energy flux in the 0.5-8 keV band is 
$\sim 2.3 \times  10^{-11}$ 
${\rm~(erg~cm^{-2}~s^{-1}})/({\rm counts~s^{-1}})$ for a source 
with a power law of a photon index $\sim 2$
and a total X-ray-absorbing gas column density of $\sim 1 \times 10^{22} 
{\rm~cm^{-2}}$, reasonable approximations for sources in \xs\ (see \S 4.1).

\subsection{\xs\ X-1}

\begin{figure}[!htb]
 \centerline{
      \epsfig{figure=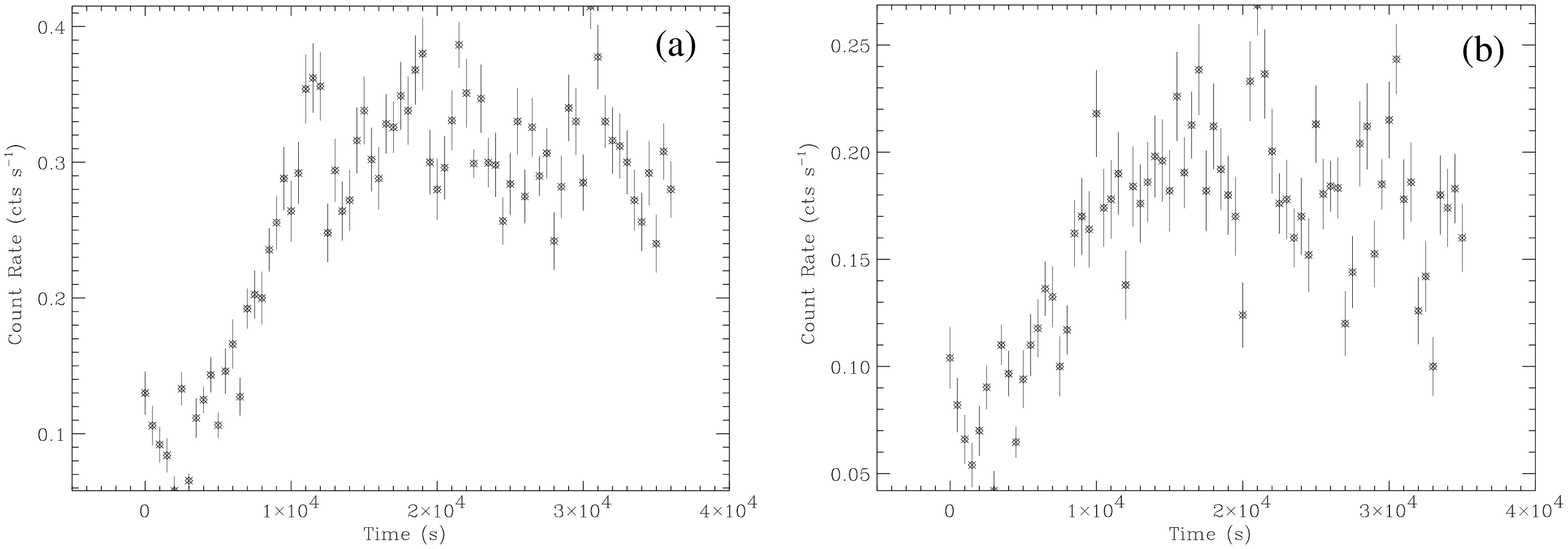,width=1\textwidth,angle=0}
    }
\caption{\small Variability of X-1 in the PN 0.5-2.0 keV (a) and 
2.0-7.5 keV (b) bands.  The events are shown in 500 s bins, with corrections for partially binning on Good Time Interval boundaries, accepting only bins with 
exposure greater than 25\% of the bin width. }
    \label{fig:var}
\end{figure}

The \xmm\ observations show a large intensity variation of X-1,
a factor of up to $\sim 6$ and 4 in 
the PN 0.5-2.0 keV  and 
2.0-7.5 keV bands and on a time scale of $10^4$ s 
(Fig.~\ref{fig:var}). This amplitude is a factor of $\sim 2$ greater 
than that during the \chandra\
observation \citep{Bauer04}. Similar variability is also seen in the MOS data. 
The large luminosity dip, as seen in the light curves, could be
due to an eclipsing of the accreting compact source by its companion.  
We detect no apparent periodic signal; both the FFT power 
spectrum and the period folding analysis show too much confusion from the harmonics
and their aliases of the CCD readout period (73.4 ms). 

\begin{figure}[!htb]
  \centerline{
      \epsfig{figure=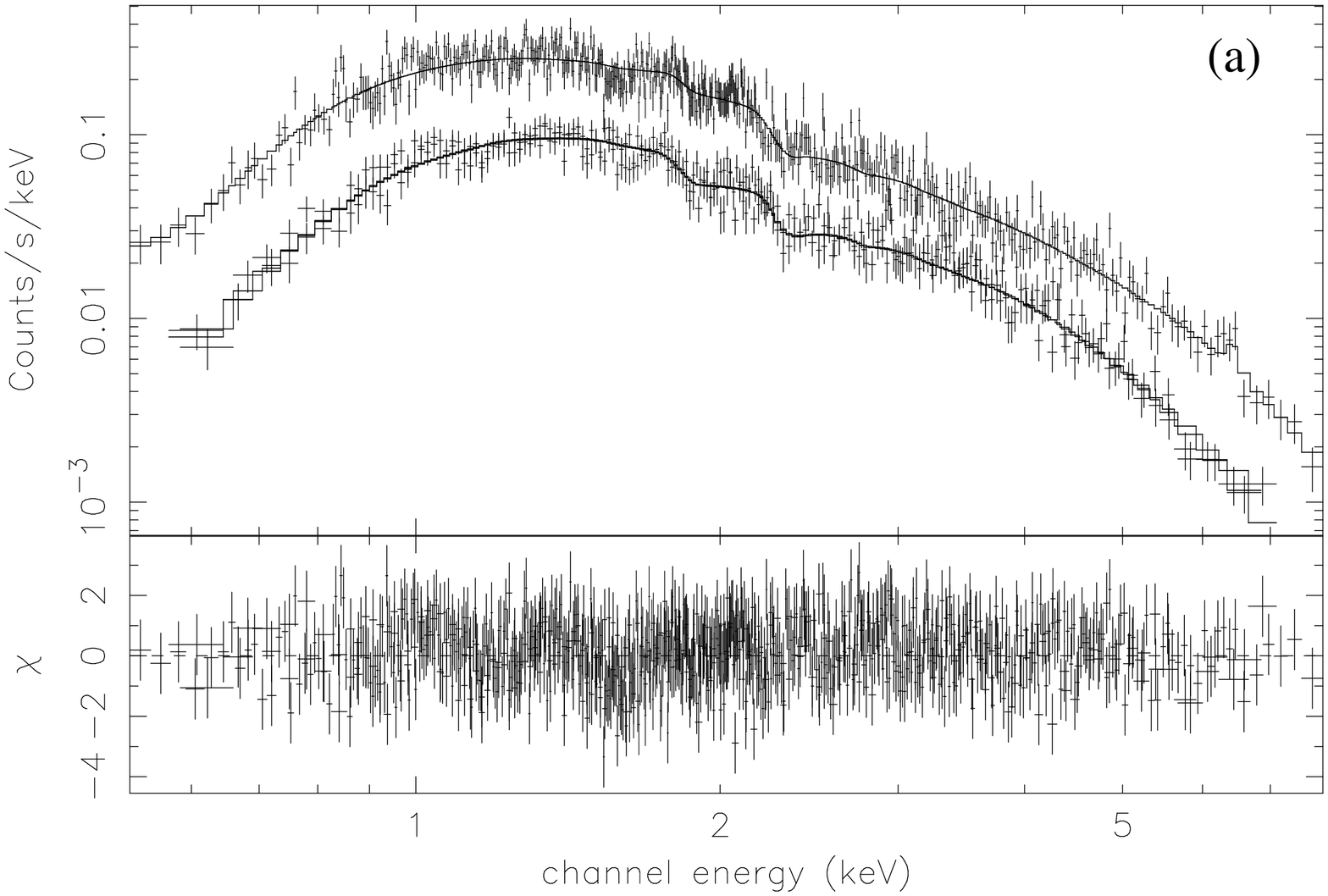,width=0.5\textwidth,angle=0}
      \epsfig{figure=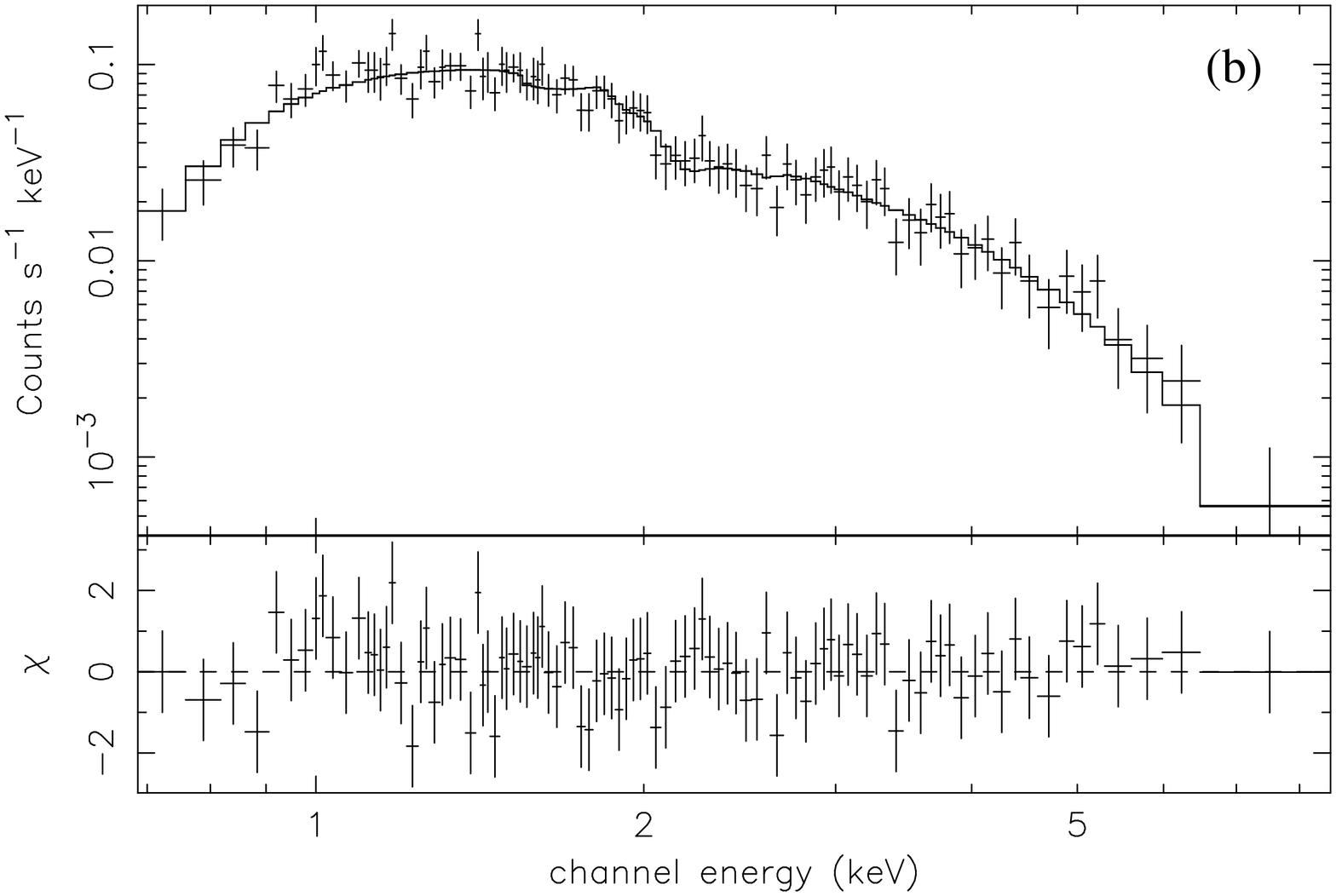,width=0.5\textwidth,angle=0}
    }
  \caption{The PN+MOS (a) and ACIS-S (b) spectra of \xs\ X-1. The histograms show
the best-fit CMCD model.
    \label{fig:spec_x1}}
\end{figure}

Fig.~\ref{fig:spec_x1}a shows that the \xmm\ spectra of X-1 are 
nearly featureless (except for various instrumental 
features). We 
fit the spectra with various models that are typically 
used for X-ray binaries, 
together with a foreground photoelectric absorption. The 
overall metal abundance of the absorbing gas is allowed to vary, relative
to hydrogen, although the relative metal abundances are fixed to the pattern 
of Balucinska-Church \& McCammon (1992). We find that a simple 
power law model with a best-fit $\chi^2/dof = 1126/812$
can be ruled out with a null hypothesis probability of $10^{-12}$;
the spectra appear to have an intrinsically blackbody-like convex shape. 
But the spectra are reasonably well characterized ($\chi^2/dof = 863/812$) 
by the  multi-color disk model (MCD; i.e.,
{\em diskbb} in the {\em XSPEC} spectral analysis 
software package; Arnaud 1996; 
Makishima et al. 1986). The fitted inner disk 
temperature and radius are $T_{in} = 1.18 (1.16-1.20)$ keV and $R_{in} = 
15 (14 - 16) /({\rm cos} \theta)^{1/2}$ km, where $\theta$ is the inclination 
angle of the disk, while the absorption column 
density is $N_H = 1.25 (1.20-1.29) 
\times 10^{22} {\rm~cm^{-2}}$ with a metal abundance $< 0.01$ solar;
a fit with the fixed solar metal abundances is not acceptable 
($\chi^2/dof = 1045/813$). This unusually low metal abundance is probably
not physical and may be due to uncertainties in the spectral
calibration at low energies.
The absorption-corrected luminosity of the source is $1.2 
\times 10^{38} {\rm~erg~s^{-1}}$ in the 0.3-8.0 keV band. 
Both the spectral shape and the luminosity of X-1 are very similar to 
those of the well-studied persistent X-ray 
binaries with stellar mass BHs (i.e., LMC X-1, LMC X-3, and Cyg X-1).

To further characterize the putative BH as well as the accretion disk of
X-1, we fit the \xmm\ spectra with a  self-consistent Comptonized MCD 
(CMCD) model (Wang et al. 2004; Yao \etal 2005). This model is implemented
in XSPEC as a table of spectra generated from Monte-Carlo simulations of 
the Comptonization, which assumes a spherically symmetric, thermal corona
around an accretion disk. The fitted parameters of the CMCD model are the 
corona optical depth $\tau = 1.7 (1.1 - 2.4)$, effective 
radius  $R_c = 50 (43-73)$ km, electron temperature $\lesssim 15$ keV, 
and disk 
inclination angle $\theta 
\lesssim 57^\circ$, as well as
$T_{in} = 1.11 (1.06 -  1.17) $ keV and $R_{in} = 25(19 - 29)$ km. 
The obtained $N_H$ is almost the same as that obtained for the MCD model.
The fit is satisfactory  ($\chi^2/dof = 837/809$). 
Following Wang et al. (2004), we further infer the mass of
the BH by assuming that $R_{in}$ represents the 
radius of the last stable circular orbit, after correcting for various general 
relativity spectral hardening and Doppler shift effects 
(a factor of $\sim 2$ total correction). 
For a non-spinning BH, our inferred mass is 
4.1(3.1, 4.9) M$_\odot$, which depends weakly on $\theta$
(within its 90\% uncertainty range). But for an extreme spinning BH, the mass
could be a factor of up to 6 greater, sensitively depending on $\theta$
(see Fig.~2 in Wang et al. 2004). 

There is a marginal presence of an emission line at 
$\sim 6.4$ keV, presumably representing the K$\alpha$
transition of neutral or weakly ionized iron.
The inclusion of a narrow Gaussian line with the centroid fixed 
at this energy  leads to a small 
improvement of the fit ($\chi^2/dof = 829/808$) and does not significantly change
the parameters of the other spectral components. The 
luminosity of the line is $\sim 2.8 (0 - 5.8) \times 10^{35} 
{\rm~ergs~s^{-1}}$,
corresponding to an equivalent width of $\sim 70 $ eV.

To test the spectral variability, we further extract two separate sets of spectra
in the time intervals less and greater than 8000 s in Fig.~\ref{fig:var}.
However, the total of PN+MOS counts in the lower flux interval is only about 470.
The two spectra show no statistically significant difference.

For comparison with the \xmm\ spectra, we further extract a 
\chandra\ ACIS-S spectrum of X-1  from a circle
of 8\as radius and a local background spectrum from a concentric 
annulus of radius 32\as-67\as. We use the newly improved
CIAO routine {\sl mkacisrmf} to generate the spectral response matrices
and include Davis's model for 
the pile-up correction (Davis 2001). 
Our obtained pile-up fraction is similar to that found in Bauer \& Brandt (2004,
who use a different method entirely).

The same CMCD model without any parameter adjustment 
gives an acceptable fit to this ACIS-S 
spectrum of X-1 ($\chi^2/dof = 77/97$); there is only a slight excess of
the observed soft X-ray flux above the model. Simply letting one parameter
free [e.g., $\tau =0.96 (0.69 - 1.3)$] improves the fit
[$\chi^2/dof = 62/96$; F-test probability for no improvement is
$5 \times 10^{-6}$;
Fig.~\ref{fig:spec_x1}b]. The inferred 0.3-8 keV luminosity is almost exactly the same
as the mean value during the \xmm\ observation. Therefore, there is no substantial
difference in both the spectral shape and luminosity 
between the \xmm\ and \chandra\ spectra of X-1. 
Fig.~\ref{fig:spec_x1}b shows 
little evidence for residual, emission line-like features that are
apparently present in the X-1 spectrum extracted by \citet{Bauer04}, 
at somewhat unexpected energies (e.g., $\sim 2.25$ and 3.65 keV; 
their Fig.~4 and discussion). Therefore, these features of
the spectra are likely data processing artifacts.

\begin{figure}[!thb]
  \centerline{
      \epsfig{figure=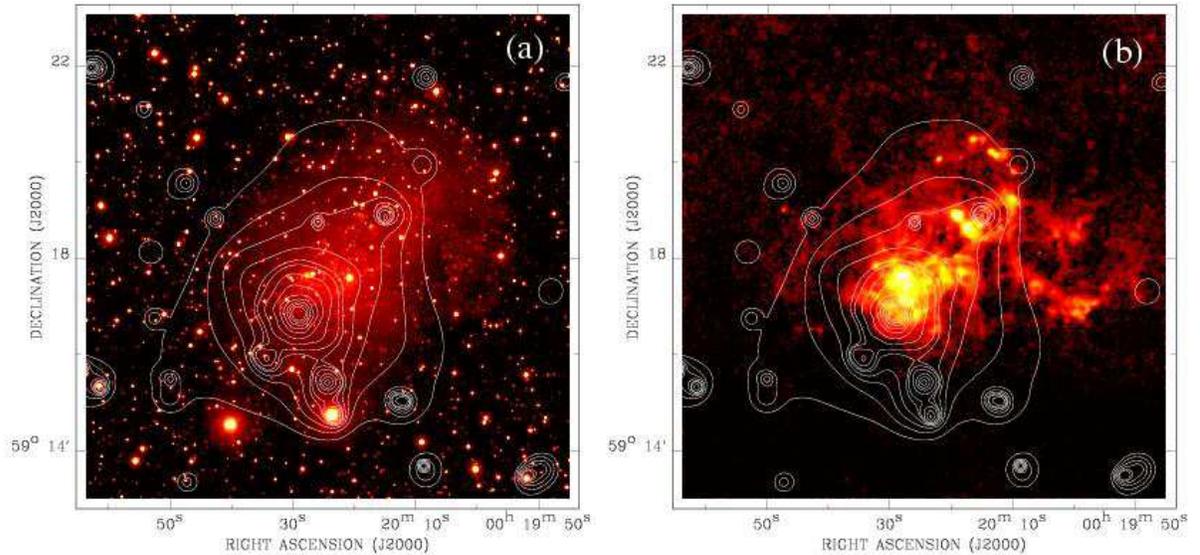,width=1.\textwidth,angle=0}
    }
  \caption{EPIC 0.5-2 keV intensity contours overlaid on (a) optical R-band 
and (b) H$\alpha$ images of \xs\ \citep{Gil03}. 
The X-ray data have been adaptively smoothed
with the CIAO routine {\sl csmooth} with a signal-to-noise ratio of $\sim 3$. 
The contours levels are at
1.1,      1.8,      2.9,      4.4,      6.2,      7.3,     11,     22,     44,     88,    183,    366, and
    731 $\times 10^{-3} {\rm~cts~s^{-1}~arcmin^{-2}}$.
    \label{fig:xc_c}}
\end{figure}

\begin{figure}[!htb]
  \centerline{
      \epsfig{figure=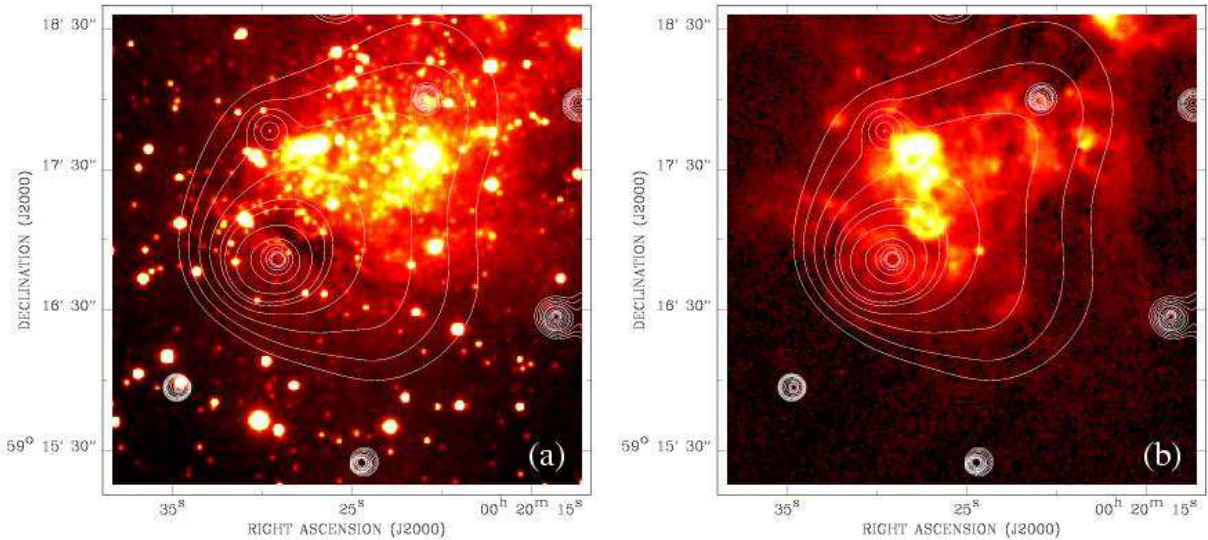,width=1.\textwidth,angle=0}
    }
  \caption{Close-up of the regions covered by the ACIS-S data. While the images
are smoothed in the same way as in Fig.~\ref{fig:xc_c}, the ACIS-S 
 0.7-4.5 keV intensity contours 
are at 3.9, 4.3, 4.9, 5.7, 6.7, 7.3, 9.3, 15, 27,  50, 100, 
200, and 400 $\times 10^{-3} {\rm~cts~s^{-1}~arcmin^{-2}}$.
    \label{fig:acis_xc_c}}
\end{figure}

\subsection{Diffuse X-ray emission}

There are clear indications for
the presence of diffuse soft X-ray emission from \xs. Fig.~\ref{fig:xc_c}
shows the large-scale soft X-ray intensity distribution of the field.
The distribution is calculated after a subtraction of the blank-sky background,
which is normalized to the local source-removed 
intensity estimated in an annulus of off-axis radius $8^\prime-10\farcm5$ 
in the \xs\ PN observation. In this step, we can 
approximately account for the background variation 
from one observation to another, particularly important for \xs\  because 
of its low Galactic latitude position.  The soft X-ray
structure near X-1 can only be examined in the high resolution \chandra\
observation (Fig.~\ref{fig:acis_xc_c}; see also \citet{Bauer04}). 
To quantify the diffuse soft X-ray component, one needs to excise the
detected sources and to estimate their residual contamination as well
as sources below our detection limit. 
With the limited spatial resolution of the \xmm\ data, a clean separation
of the sources from the diffuse component is difficult.
One might use a large EER to exclude a large
fraction of source counts. However, the 90\% EER at 1.5 keV, 
for example, is $\sim 50$\as on-axis and increases to $\sim 90$\as at an 
off-axis angle of 12$^\prime$. The adoption of such an EER for all the sources
would leave regions too sparse for a diffuse emission study. 
A smaller EER is adequate for the bulk of faint sources (CR $\lesssim 
0.01 {\rm~counts~s^{-1}}$), which
would leave acceptable amounts of contamination. 
For such a source, our adopted 
removal radius is twice the source detection radius for both
the ACIS-S and PN data (or $\sim 75\%$ PSF EER for the PN data). 
For brighter sources, we scale this radius by a factor of 
$1+{\rm log}(CR/0.01)$. The adopted source removal regions are 
outlined in Figs.~\ref{fig:sou_pn} and \ref{fig:sou_acis}.

\begin{figure}[!htb]
  \centerline{
      \epsfig{figure=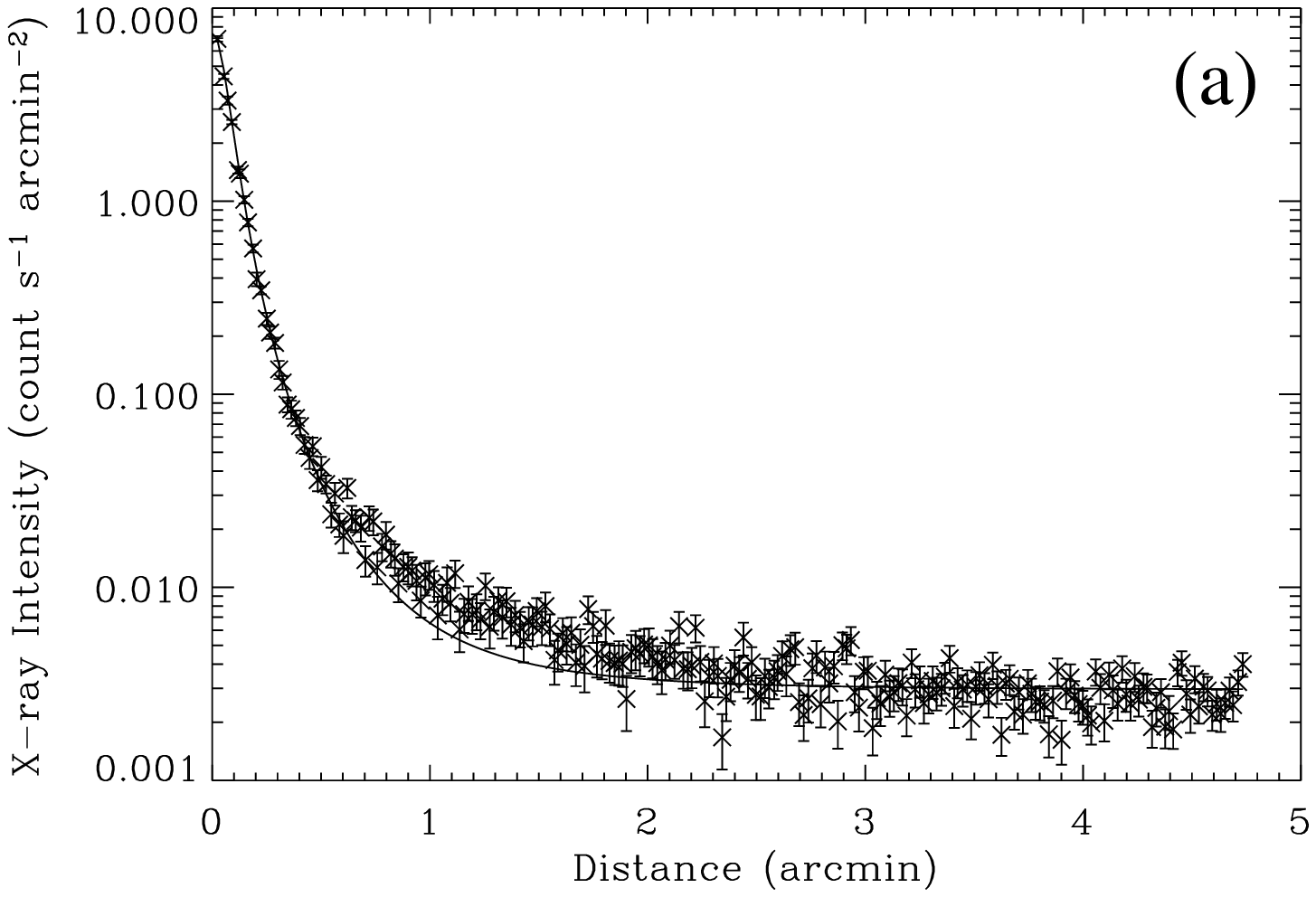,width=0.5\textwidth,angle=0}
      \epsfig{figure=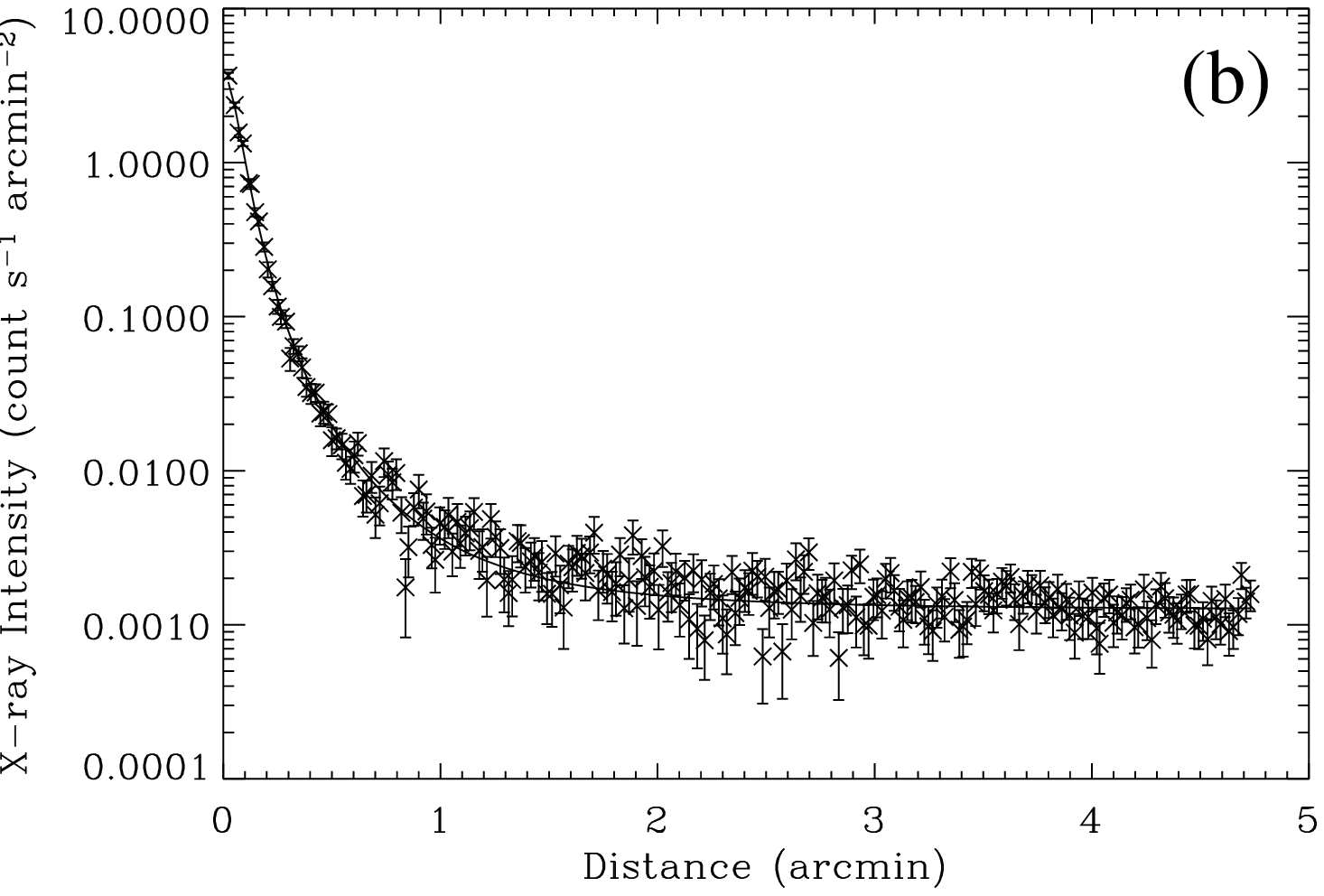,width=0.5\textwidth,angle=0}
    }
  \caption{Radial surface intensity profiles around \xs\ X-1 
in the PN 0.5-2 keV (a) 
and 2-4.5 keV bands (b). Background is not subtracted from the data, 
and is included in the PSF fits, which are shown as the solid lines.
    \label{fig:x1_rbp}}
\end{figure}

Of course, the remaining ``diffuse'' enhancement is still contaminated
by the PSF wing of X-1 and other discrete sources. The on-axis PN PSF
can be characterized by a King function,
\begin{equation}
   I_{PSF}(R)={A}\left[1+\left(\frac{R}{r_0}\right)^2\right]^{-\alpha},
\end{equation}
where $R$ is the off-source angular distance and $A$ is a normalization factor, 
while the core radius 
${r_0} \approx 5\farcs5$ or 5\farcs2 and index $\alpha \approx 1.6$ or 1.5 for the
the 0.5-2 keV or 2-4.5 keV bands, respectively\footnote{http://xmm.vilspa.esa.es/external/xmm\_user\_support/documentation/uhb/node16.html}.
We compare the PSF with 
 the radial profiles of the source-removed PN intensity around \xs\ 
X-1 by fitting the normalization and local uniform background 
(Fig.~\ref{fig:x1_rbp}). The PSF gives a 
reasonably good fit to the profile
in the 2-4.5 keV band  ($\chi^2/dof = 269.8/237$). However, the fit in
the 0.5-2 keV band is not acceptable  ($\chi^2/dof = 522.9/237$);
an excess of the observed
intensity above the PSF is apparent at the annulus of 
the off-source distance between
$\sim 0\farcm6 - 2\farcm4$.
The total net CR in this annulus is $\sim 0.022 $ (0.005) ${\rm~counts~s^{-1}}$ in the 
0.5-2 keV (2-4.5 keV) band. We estimate that $\sim 0.014 {\rm~counts~s^{-1}}$ in the 0.5-2 keV 
band is due to the residual point-like contribution, assuming
that it dominates the 2-4.5 keV band CR and has a mean spectrum similar to the 
best-fit absorbed power law of the \chandra\ detected sources (\S 3.1). The 
remaining  $\sim 0.008 {\rm~counts~s^{-1}}$ in the 0.5-2 keV band 
may be a truly diffuse component. The luminosity of contribution is about the
same as that of the detected sources within \xs, excluding X-1 (\S 3.1). 
The faint point source contribution
is probably a bit overestimated here, if it is less contaminated by 
the Galactic stellar population (which tends to contribute more soft X-rays) 
than the detected source spectrum.

\begin{figure}[!htb]
  \centerline{
      \epsfig{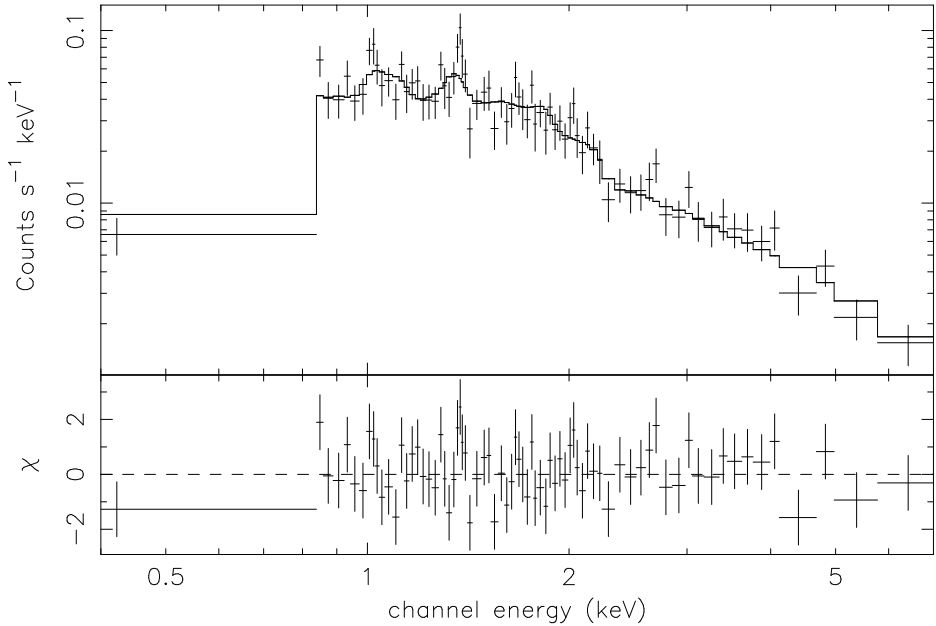}
    }
  \caption{PN spectrum of the diffuse X-ray enhancement of \xs,
together with the fitted thermal plasma plus power law model. 
    \label{fig:spec_d}}
\end{figure}

We may also decompose the point-like and diffuse components in the PN data
spectrally. We extract a spectrum
of the diffuse component in the $0\farcm6 - 2\farcm4$ annulus.
From the same region, we also obtain a background spectrum from the 
re-projected blank-sky data. The use of this background spectrum minimizes the
effect of instrumental background variation across the detector.
However, the sky background typically varies from 
one field to another. To account
for this variation, we adjust the background spectrum 
by adding the spectral difference 
between the \xs\ and blank-sky data in an annulus 
between $5\farcm0 - 9\farcm3$ off X-1.
In all of these spectral extractions, the source regions are removed from both
the \xs\ and the blank-sky data. Fig.~\ref{fig:spec_d} shows 
the background-subtracted 
net spectrum of the diffuse X-ray component of \xs. 
The apparent emission line features at $\sim 1.01$ and $1.35$, 
corresponding to the Ne X and Mg XI K$\alpha$ transitions, indicates
that part of the component has a thermal origin.
Therefore, we fit this spectrum
with the combination of an optically thin thermal 
plasma (XSPEC {\sl MEKAL} model) and
a power law to characterize the residual point-like source contribution. 
The fit
is satisfactory ($\chi^2/dof = 64.6/67$; 
Fig.~\ref{fig:spec_d}) and gives reasonable model parameters: $kT= 0.30
(0.20 - 0.50)$, metal abundance $=0.84 (>0.11)$, power law photon
index $= 2.3(2.0-2.5)$, and $N_H = 1.1 (0.61-1.43) \times 10^{22} 
{\rm~cm^{-2}}$ (assuming the  solar
metal abundance for the absorbing gas). The index is 
marginally greater than that inferred from the fit to the accumulated
\chandra\ source spectrum (\S 3.1), which is probably due, in part,
to our demand for the same $N_H$ for both the power law 
and the plasma components. Indeed, these two power laws (one from
the source spectrum and the other from the spectral 
decomposition here) give comparable predictions of the relative contribution
of point-like sources
in the 0.5-2 keV and 2-4.5 keV bands. Thus we conclude that the residual point-like source
contribution,
corrected for the absorption,  is $\sim 2\times 10^{37} {\rm~erg~s^{-1}}$
in the 0.5-2 keV band and the corresponding hot gas luminosity is a factor of $\sim 4$ higher. 
If an extrapolation of the best-fit thermal plasma model to 
energies outside the band
is a good approximation, the total cooling
rate of the plasma should then be about $2 \times  10^{38} {\rm~erg~s^{-1}}$.
 
\section{Discussion}

\subsection{Discrete X-ray sources}

Following Wang (2004a), we estimate the
 contribution of background AGNs in the sources detected from the PN data.
This estimate uses a  X-ray source log $N$ - log $S$ relation in the 
0.5-2 keV band from the \chandra\ deep survey. We assume that these AGNs
have a power law with a typical photon index of $\sim 1.7$ and are subject to
an average foreground absorption of the Galactic \ion{H}{1} column density 
$\sim 5 \times 10^{21} {\rm~cm^{-2}}$. Accounting for both the 
sensitivity incompleteness and
the Eddington bias of our source detection in the 0.5-7.5 keV band, we
estimate the AGN contribution is $\sim 20$. If the absorption is higher
(e.g., including the column contribution from \xs), the contribution
would be smaller.  Although this estimate is 
quite uncertain, both statistically and systematically (Wang 2004), it is 
clear that the bulk of our detected PN sources are either Galactic or
associated with \xs.

The identification of individual X-ray sources in the \xs\ field 
is generally not easy, because
of the galaxy's low Galactic latitude position ($b = -3\fdg3$). 
In addition to the Galactic \ion{H}{1} column density, there is evidence for
substantial contributions from molecular gas in the Galaxy and/or
gas inside \xs. The comparison between the radio and H$\alpha$ fluxes 
of \ion{H}{2} regions, for example, 
indicates $E_{B-V} \sim 2$  (Yang \& Skillman 1993). From the Galactic 
relation 
$N_H/E_{B-V} = 4.8  \times 10^{21} 
{\rm~cm^{-2}}$ (Bohlin, Savage, \& Drake 1978), we 
infer $N_H \sim 1 \times 10^{22} {\rm~cm^{-2}}$,
which is quite similar to the X-ray-absorbing
gas column density that we have estimated from the above spectral 
analysis of X-1 and the sources in the main body of \xs. 

Considering the general high extinction expected for the objects in the field,
we first use the Two Micron All Sky Survey (2MASS) 
All-Sky Catalog of Point Sources \citep{Cutri03} to search for potential 
counterparts. We cross-correlate the spatial positions of the objects in the catalog 
with those from Tables~\ref{pn_source_list} and \ref{acis_source_list}, using 
a matching radius of 4\as\ for PN sources and 2\as\ for ACIS sources.
The radius is chosen to be greater than the $1\sigma$ statistical position
uncertainty of almost all the sources. Table~\ref{tab_sou_id} presents
the matching results, including the position offset of each match; 
there is no match with multiple 2MASS objects. The table also includes 
the J, H, and K$_s$ magnitudes of the matched 2MASS objects; the 
$3\sigma$ limiting sensitivities of the catalog are 17.1, 16.4 and 15.3 
mag in the three bands. The X-ray sources with the 2MASS matches 
tend to have soft X-ray spectral characteristics, having negative 
hardness ratios  and/or being detected preferentially in the S band 
(Tables~\ref{pn_source_list} and \ref{acis_source_list}). These 2MASS
objects, many of which also show up in optical images, 
should mostly be Galactic stars. The 2MASS matches with those few sources 
with relatively hard X-ray spectral characteristics (namely XP-16, XP-34, and XA-11) are
all very faint (e.g. K$_s \gtrsim 14.6$); they may be stars in distant parts
of the Galaxy. Some of the matches in Table~\ref{tab_sou_id}
should represent chance projections of 2MASS objects within the 
matching regions. We estimate the expected 
number of such chance projections to be 0.25 for
the ACIS sources and 2.6 for the PN sources, based on 
the surface number density of 2MASS objects within annuli
of radius 4\as-15\as around the X-ray sources. 

We next use the NASA/IPAC Extragalactic Database and the SIMBAD Astronomical
Database to search for identified objects within 2\as
of each X-ray source position.  We find that XP-29/XA-9
 matches with an unresolved nonthermal radio source (Yang \& Skillman 1993; 
Chyzy et al. 2003), XP-9 with HD 1486 --- an
eclipsing binary of Algol type, and XP-25 with HL90 --- 
an HII region in \xs. Increasing the search radius to 4\as only leads to
one additional correspondence, i.e., XP-57 with the NRAO VLA sky survey 
radio source NVSS J002108+591132. 

\subsection{\xs\ X-1}

We have shown that X-1 has a blackbody-like X-ray spectrum. The inferred 
inner disk temperature ($T_{in} \sim 1.1$ keV) is consistent with those found in 
LMC X-1, LMC X-3, and Cyg X-1 --- stellar mass BH X-ray binaries. In contrast,
X-ray binaries with intermediate mass BH candidates have substantially 
lower $T_{in}$ values ($\sim 0.1-0.3$ keV; e.g., Wang et al. 2004).  
The spectrum of an accreting neutron star
in an X-ray binary system is typically  dominated by  a
power law component and/or an optically thin X-ray photo-ionized plasma (e.g.,
Paerels et al. 2000). Therefore, \xs\ X-1 is most likely 
a stellar mass BH associated with the W-R star [MAC92] 17A.

The only other known candidate for such an X-ray luminous BH/W-R binary 
system is Cyg X-3
at a distance of $\sim 9$ kpc in our Galaxy ($l, b = 79\fdg8, 0\fdg$7), 
although  the true nature of this source remains uncertain
(e.g., Mitra 1998). The X-ray
spectrum of Cyg X-3 typically shows strong lines, which 
are interpreted as the evidence for a tenuous X-ray photo-ionized plasma, 
presumably in the stellar wind from the W-R companion star 
(Liedahl \& Paerels 1996). But occasionally, the line emission weakens 
substantially, probably during a very high luminosity state 
(Smale et al. 1993). The spectrum in this state is very similar
to that of \xs\ X-1 in terms of both $T_{in}$ and the
X-ray luminosity. Longer observations of \xs\ X-1 
are needed to further the comparative study (e.g., to 
capture the low luminosity state
of the source). Despite its projected location near the Galactic plane, 
\xs\ X-1 is still in a line of sight that is considerabally less confusing
and obscured than that of Cyg X-3. Therefore, the study of \xs\ X-1 may provide 
new insights into the end products of most massive stars (e.g., 
the progenitor mass required to form black holes in close binaries;
e.g., Wellstein \& Langer 1999; Clark \& Crowther 2004).

\subsection{Diffuse X-ray emission}

Evidence for the diffuse soft X-ray emission in the immediate vicinity of \xs\ X-1 
has also been presented by \citet{Bauer04}. From a comparison of the 
radial intensity profile around the source with the PSF of the \chandra, they
find that a diffuse emission enhancement with a 
total ACIS-S CR of $\sim 1 \times 10^{-3} {\rm~counts~s^{-1}}$ 
extends $\sim 17$\as, a scale comparable to the radius of the 
nonthermal radio shell discovered by Yang \& Skillman (1993). 
This result is not significantly affected by our removal of the CCD-readout streak,
which contributes $\sim 16\%$ of the enhancement.  But its morphology
is changed from the slightly northeast-southwest extension before the removal
(see Fig.~1 in \citet{Bauer04}) to the northwest orientation after
the removal (Fig.~\ref{fig:acis_st}). 

We have further shown that the lower surface brightness 
diffuse soft X-ray emission 
extends at least $\sim 2^\prime$ (Fig.~\ref{fig:xc_c}), and 
possibly as far as $\sim 4^\prime$ (Fig.~\ref{fig:acis_xc_c}), 
northwest from X-1. The 
morphology of the emission seems to represent an extension of 
recent intense star-forming
regions where the bulk of known W-R stars are located. The overall extent
of the emission appears to outline the boundaries of prominent 
H$\alpha$ streamers/features; as illustrated by \citet{Wil98} 
with a high-quality 
H$\alpha$ image, these features correspond to the Shells 6 (located slightly
southwest to
X-1), 7 (northwest to X-1), 8 (around X-1), and possibly 4/5 further to the
northwest. This morphological similarity 
suggests an association of the diffuse soft X-ray emission with the 
H$\alpha$ streamers, i.e., a scaled-up version of the 30 Doradus Nebula
in the Large Magellanic cloud \citep{wang99}.

The soft spectral characteristics of the diffuse X-ray emission 
indicates that it represents diffuse hot gas in \xs. Our spectral
analysis also shows marginal evidence for enhanced metal abundances.
 But with a temperature of a few $10^6$ K, as we have 
inferred, this hot gas cannot be confined by the gravity of \xs. 
Without sufficient cooling,
the gas is likely to escape. Indeed, as in many other active 
star-forming galaxies
(Wang et al. 2001), there is an apparent ``missing'' energy 
problem (Wang 2004b).
With a typical mass-loss rate of $\sim 10^{-5} {\rm~M_\odot~yr^{-1}}$ 
and a wind speed of $\sim 10^{3} {\rm~km~s^{-1}}$, a W-R star is 
expected to have  a mechanical energy
luminosity of $\sim 3 \times 10^{36} {\rm~erg~s^{-1}}$.
The W-R star population (\S 1) alone could provide sufficient energy to
balance the cooling of the diffuse soft X-ray-emitting gas 
($2 \times  10^{38} {\rm~erg~s^{-1}}$; \S 3.3). Of course, 
stellar winds from other massive stars as well as supernova blastwaves together
should supply substantially more energy. But this energy is not observed
in X-ray radiation. Most likely, the energy is used to drive the expansion of 
the various observed energetic structures around recent massive star forming
regions. This process is probably happening in
the galaxy's southeast portion, where the hot gas is apparently still 
confined, naturally explaining the overall morphological
similarity between the diffuse soft X-ray emission and the 
H$\alpha$ streamer/features. Once the hot gas is out of the
confinement of the ISM, the X-ray radiative cooling becomes even less efficient. 
This transition has probably occurred around previous star formation 
regions in the western portion of the galaxy. These regions, traced by 
large (probably quiescent) faint H$\alpha$ shells 
\citep{Thurow05, Wil98}, show little diffuse X-ray enhancement.

\section{Summary}

We have conducted a careful analysis of the \xmm\  and \chandra\ X-ray 
CCD observations of 
the nearby starburst galaxy \xs. In particular, we have devised an 
effective method to remove the CCD-readout streaks of the bright 
source \xs\ X-1. The main results of our analysis are as follows:
\begin{itemize}

\item A list  of 73 \xmm\ and 28 \chandra\ detections of point-like sources 
in the \xs\ field is presented, including preliminary multi-wavelength 
identifications based on existing databases. While a large portion of these
sources are likely to be foreground and background objects, those
associated with \xs\ tend to show hard X-ray spectral
characteristics consistent with high-mass X-ray binaries.

\item We confirm that X-1 likely represents a BH/W-R binary system.
The mean X-1 luminosity of $\sim 1.2 \times 10^{38} {\rm~erg~s^{-1}}$
in the  0.3-8.0 keV band is about the same 
during the \xmm\ and \chandra\ observations. But the source shows 
large luminosity variation by a factor of  up to 
$\sim 6$ on time scales of $\sim 10^4$ s.
The spectrum of X-1 is well characterized by a Comptonized
multi-color blackbody accretion disk with an inner disk 
temperature $T_{in} \approx 1.1$ keV, typical of those  
persistent X-ray binaries with stellar mass BHs (i.e., LMC X-1, 
LMC X-3, and Cyg X-1). For X-1, we infer the mass of
the putative BH as $\sim 4$ M$_\odot$ if it is not spinning, or 
a factor of up to $\sim 6$ higher if the spinning is important.

\item The presence of the diffuse soft X-ray emission in \xs\ is revealed.
The emission is morphologically oriented along the optical main
body and is greatly enhanced around most active star-forming regions 
of the galaxy. The extent of the emission resembles that of 
luminous H$\alpha$ streamers, emanating from starburst regions.
The diffuse soft X-ray emission most likely arises from chemically-enriched
hot gas with a characteristic temperature of $\sim 0.3
$ keV and a total 0.5-2 keV luminosity of 
$\sim 8 \times 10^{37} {\rm~erg~s^{-1}}$. Although its total cooling
rate ($\sim 2 \times 10^{38} {\rm~erg~s^{-1}}$) is still 
quite uncertain, this hot gas could be easily maintained by 
the mechanical energy input from the observed large W-R star population 
in the galaxy.
But the X-ray radiation is unlikely to be the main cooling mechanism of the hot gas,
which apparently powers various energetic ISM features
observed around the starburst regions and will eventually escape from
the galaxy.

\end{itemize}

These results demonstrate that \xs\ is a high-energy power house,
hosting numerous energetic stellar and interstellar activities, which are most likely
the end products of massive stars. 

\acknowledgements We thank the referee Franz Bauer for valuable comments on the work,
which is supported by NASA through the grant NAG5-13582 and the Massachusetts Space Grant.
This publication makes use of data products from the Two Micron All Sky Survey, which is a joint project of the University of Massachusetts and the Infrared Processing and Analysis Center/California Institute of Technology, funded by the National Aeronautics and Space Administration and the National Science Foundation.
  \vfill
\eject
 
\vfil
\eject
\begin{deluxetable}{lrrrrrrrr}
  \tabletypesize{\footnotesize}
  \tablecaption{{\sl XMM-Newton} Source List \label{pn_source_list}}
  \tablewidth{0pt}
  \tablehead{
  \colhead{Source} &
  \colhead{XMMU Name} &
  \colhead{$\delta_x$ ($''$)} &
  \colhead{CR $({\rm~cts~ks}^{-1})$} &
  \colhead{HR} &
  \colhead{HR1} &
  \colhead{Flag} \\
  \noalign{\smallskip}
  \colhead{(1)} &
  \colhead{(2)} &
  \colhead{(3)} &
  \colhead{(4)} &
  \colhead{(5)} &
  \colhead{(6)} &
  \colhead{(7)} 
  }
  \startdata
   1 &  J001834.79+591946.8 &  1.8 &$    19.28  \pm   2.23$& --& $-0.38\pm0.10$ & B, S \\
   2 &  J001845.81+592331.9 &  3.1 &$     5.57  \pm   1.44$& --& $-0.72\pm0.18$ & S, B \\
   3 &  J001855.81+591240.1 &  1.5 &$    19.06  \pm   1.93$& --& $-0.38\pm0.09$ & S, B \\
   4 &  J001857.44+590458.5 &  3.4 &$     6.85  \pm   1.67$& --& --& B \\
   5 &  J001902.38+591623.3 &  2.2 &$     6.74  \pm   1.12$& --& $ 0.70\pm0.15$ & B, S \\
   6 &  J001909.72+591916.0 &  2.7 &$     4.74  \pm   1.00$& --& --& B, S \\
   7 &  J001910.26+592455.9 &  3.7 &$     3.76  \pm   1.00$& --& --& B, S \\
   8 &  J001917.94+591231.9 &  2.2 &$     6.89  \pm   1.11$& --& $-0.28\pm0.16$ & B, S \\
   9 &  J001918.64+590820.7 &  0.3 &$   529.36  \pm   9.60$& $-0.72\pm0.02$ & $-0.11\pm0.02$ & B, S, H \\
  10 &  J001922.60+591541.7 &  3.3 &$     2.85  \pm   0.74$& --& --& B \\
  11 &  J001925.54+591417.5 &  2.1 &$     5.34  \pm   0.92$& --& $-0.39\pm0.15$ & S, B \\
  12 &  J001926.58+591103.5 &  3.2 &$     3.12  \pm   0.85$& --& --& S, B \\
  13 &  J001927.16+592423.6 &  3.1 &$     4.30  \pm   0.98$& --& --& B, H \\
  14 &  J001929.33+590420.9 &  3.0 &$     7.93  \pm   1.64$& --& $ 0.07\pm0.19$ & B, S \\
  15 &  J001938.11+591652.6 &  2.9 &$     3.02  \pm   0.64$& --& --& B, S \\
  16 &  J001944.06+591815.9 &  2.5 &$     3.33  \pm   0.66$& $ 0.47\pm0.18$ & --& H, B \\
  17 &  J001947.84+591711.5 &  2.0 &$     3.74  \pm   0.68$& $ 0.25\pm0.19$ & --& B, H \\
  18 &  J001949.48+590620.6 &  3.0 &$     6.24  \pm   1.26$& --& --& B, S \\
  19 &  J001949.69+591334.3 &  2.5 &$     3.10  \pm   0.66$& --& --& B \\
  20 &  J001950.60+590440.4 &  3.5 &$     5.66  \pm   1.32$& --& --& B, S \\
  21 &  J001951.82+591328.5 &  2.2 &$     3.15  \pm   0.64$& --& $-0.64\pm0.16$ & S \\
  22 &  J001952.36+590704.5 &  3.3 &$     4.29  \pm   1.00$& --& --& B, S \\
  23 &  J001952.73+590854.1 &  2.7 &$     5.62  \pm   1.01$& $ 0.49\pm0.19$ & --& B, H \\
  24 &  J002008.34+592144.5 &  2.0 &$     4.72  \pm   0.73$& $ 0.35\pm0.17$ & --& B, H, S \\
  25 &  J002008.36+591953.2 &  2.5 &$     2.11  \pm   0.51$& --& --& B \\
  26 &  J002008.84+591338.5 &  2.1 &$     4.03  \pm   0.67$& --& $ 0.58\pm0.17$ & B, S \\
  27 &  J002010.78+592547.2 &  2.0 &$     6.61  \pm   1.06$& $ 0.29\pm0.18$ & --& B, H, S \\
  28 &  J002012.48+591500.4 &  1.1 &$     7.75  \pm   0.82$& $ 0.06\pm0.13$ & $ 0.66\pm0.12$ & B, S, H \\
  29 &  J002015.10+591854.0 &  1.4 &$     5.37  \pm   0.72$& --& $ 0.16\pm0.13$ & B, S \\
  30 &  J002015.84+592637.2 &  2.8 &$     4.05  \pm   0.90$& --& --& B, H \\
  31 &  J002018.03+590732.0 &  1.7 &$    10.96  \pm   1.33$& $ 0.47\pm0.12$ & $ 0.71\pm0.18$ & B, H, S \\
  32 &  J002018.31+591834.5 &  2.8 &$     1.92  \pm   0.52$& --& --& H \\
  33 &  J002021.57+591900.1 &  2.0 &$     3.52  \pm   0.63$& --& --& B, H, S \\
  34 &  J002023.16+590812.8 &  2.0 &$     6.80  \pm   1.04$& --& $ 0.40\pm0.18$ & B, S, H \\
  35 &  J002023.46+591445.5 &  1.3 &$     7.24  \pm   0.83$& --& $-0.58\pm0.10$ & S, B \\
  36 &  J002024.47+591525.7 &  0.5 &$    37.52  \pm   1.70$& $-0.03\pm0.06$ & $ 0.61\pm0.05$ & B, S, H \\
  37 &  J002025.84+591844.6 &  1.6 &$     4.15  \pm   0.67$& --& $ 0.08\pm0.20$ & B, S, H \\
  38 &  J002027.03+590613.8 &  3.0 &$     4.74  \pm   1.05$& --& --& B, S \\
  39 &  J002029.10+591651.1 &  0.1 &$   656.72  \pm   6.82$& $-0.15\pm0.01$ & $ 0.61\pm0.01$ & B, S, H \\
  40 &  J002033.97+591112.0 &  3.2 &$     1.79  \pm   0.57$& --& --& S \\
  41 &  J002034.07+591031.9 &  2.0 &$     4.13  \pm   0.72$& $ 0.73\pm0.14$ & --& B, H \\
  42 &  J002034.12+591555.6 &  1.2 &$     7.22  \pm   0.87$& --& $-0.10\pm0.13$ & B, S \\
  43 &  J002036.20+592402.6 &  2.9 &$     2.79  \pm   0.67$& --& --& B, S \\
  44 &  J002042.27+591709.6 &  2.6 &$     2.35  \pm   0.56$& --& --& B, H \\
  45 &  J002042.63+591850.3 &  2.9 &$     1.90  \pm   0.49$& --& --& B \\
  46 &  J002047.38+591935.4 &  2.1 &$     3.94  \pm   0.64$& $ 0.35\pm0.17$ & --& B, H, S \\
  47 &  J002050.24+591528.3 &  2.9 &$     2.09  \pm   0.54$& --& --& S, B \\
  48 &  J002050.26+591505.1 &  2.7 &$     2.41  \pm   0.57$& --& --& B \\
  49 &  J002052.96+591647.4 &  2.4 &$     2.54  \pm   0.55$& --& --& B, H \\
  50 &  J002053.78+592106.4 &  3.3 &$     1.52  \pm   0.51$& --& --& S \\
  51 &  J002058.62+590833.8 &  3.0 &$     3.67  \pm   0.85$& --& --& B \\
  52 &  J002100.75+591104.9 &  3.0 &$     3.28  \pm   0.73$& --& --& S, B \\
  53 &  J002101.69+591519.5 &  2.1 &$     3.66  \pm   0.68$& --& $ 0.26\pm0.16$ & S, B \\
  54 &  J002102.62+592158.0 &  1.8 &$     6.32  \pm   0.90$& $-0.00\pm0.19$ & $ 0.46\pm0.17$ & B, S, H \\
  55 &  J002103.82+590957.4 &  2.2 &$     5.62  \pm   0.94$& $ 0.87\pm0.10$ & --& H, B \\
  56 &  J002104.70+591540.2 &  1.8 &$     4.61  \pm   0.73$& --& $-0.47\pm0.13$ & S, B \\
  57 &  J002108.72+591134.4 &  2.6 &$     3.78  \pm   0.76$& --& --& B, S, H \\
  58 &  J002114.93+590844.5 &  1.7 &$    11.65  \pm   1.43$& --& $-0.33\pm0.12$ & B, S \\
  59 &  J002117.67+591839.9 &  3.4 &$     2.26  \pm   0.63$& --& --& B \\
  60 &  J002121.44+590905.7 &  0.9 &$    40.00  \pm   2.45$& --& $-0.26\pm0.06$ & B, S \\
  61 &  J002125.43+591904.2 &  1.8 &$     6.02  \pm   0.89$& --& $-0.59\pm0.13$ & B, S \\
  62 &  J002131.27+591428.0 &  2.6 &$     4.04  \pm   0.83$& --& --& B, S \\
  63 &  J002132.53+591313.6 &  3.1 &$     3.88  \pm   0.86$& --& --& B \\
  64 &  J002134.56+591443.6 &  2.8 &$     3.17  \pm   0.75$& --& --& B, S \\
  65 &  J002139.04+592243.4 &  1.9 &$     9.09  \pm   1.30$& --& $-0.32\pm0.13$ & S, B \\
  66 &  J002142.96+591207.3 &  3.4 &$     3.61  \pm   0.95$& --& --& B \\
  67 &  J002144.80+591730.8 &  3.4 &$     3.08  \pm   0.84$& --& --& B \\
  68 &  J002144.98+590736.2 &  3.3 &$     5.82  \pm   1.41$& --& --& B \\
  69 &  J002151.67+590820.1 &  3.5 &$     5.68  \pm   1.40$& --& --& B, H \\
  70 &  J002155.38+591008.5 &  2.6 &$     7.60  \pm   1.46$& --& --& B, S \\
  71 &  J002158.87+591326.7 &  1.3 &$    25.98  \pm   2.19$& --& $-0.28\pm0.08$ & B, S \\
  72 &  J002203.62+591515.4 &  3.9 &$     4.01  \pm   1.11$& --& --& B \\
  73 &  J002209.37+591357.6 &  1.7 &$    19.79  \pm   2.05$& --& $-0.09\pm0.10$ & B, S \\
\enddata
\tablecomments{The definition of the bands:
0.5--1 (S1), 1--2 (S2), 2--4.5 (H1), and 4.5--7.5~keV (H2). 
In addition, S=S1+S2, H=H1+H2, and B=S+H.
 Column (1): Generic source number. (2): 
{\sl XMM-Newton} X-ray Observatory (unregistered) source name, following the
{\sl XMM-Newton} naming convention and the IAU Recommendation for Nomenclature
(http://cdsweb.u-strasbg.fr/iau-spec.html). (3): Position 
uncertainty (1$\sigma$) calculated from the maximum likelihood centroiding.  (4): On-axis source broad-band count rate --- the sum of the 
exposure-corrected count rates in the four
bands. (5-6): The hardness ratios defined as 
${\rm HR}=({\rm H-S2})/({\rm H+S2})$, and ${\rm HR1}=({\rm S2-S1})/{\rm S}$, 
listed only for values with uncertainties less than 0.2.
(7): The label ``B'', ``S'', or ``H'' mark the band in 
which a source is detected with the most accurate position that is adopted in
Column (3). 
}
  \end{deluxetable}
  \vfill
\eject

\begin{deluxetable}{lrrrrrrrr}
  \tabletypesize{\footnotesize}
  \tablecaption{{\sl Chandra} Source List \label{acis_source_list}}
  \tablewidth{0pt}
  \tablehead{
  \colhead{Source} &
  \colhead{CXOU Name} &
  \colhead{$\delta_x$ ($''$)} &
  \colhead{CR $({\rm~cts~ks}^{-1})$} &
  \colhead{HR} &
  \colhead{HR1} &
  \colhead{Flag} \\
  \noalign{\smallskip}
  \colhead{(1)} &
  \colhead{(2)} &
  \colhead{(3)} &
  \colhead{(4)} &
  \colhead{(5)} &
  \colhead{(6)} &
  \colhead{(7)} 
  }
  \startdata
   1 &  J001954.74+591721.5 &  1.7 &$     0.40  \pm   0.21$&                                            --& --&  S, B    \\
   2 &  J002004.91+591507.4 &  0.7 &$     0.62  \pm   0.25$&                                            --& --&  B, H    \\
   3 &  J002008.41+591540.0 &  0.4 &$     0.64  \pm   0.22$&                                            --& --&  B, H    \\
   4 &  J002009.23+591818.3 &  0.9 &$     0.44  \pm   0.20$&                                            --& --&  B, H    \\
   5 &  J002011.49+591627.0 &  0.6 &$     0.32  \pm   0.17$&                                            --& --&  B, H    \\
   6 &  J002012.54+591757.8 &  0.3 &$     0.72  \pm   0.23$&                                            --& --&  H, B    \\
   7 &  J002012.75+591500.8 &  0.1 &$     5.42  \pm   0.56$&                  $ 0.27\pm0.12$ & $ 0.90\pm0.10$ &  B, S, H \\
   8 &  J002013.75+591626.6 &  0.2 &$     2.34  \pm   0.40$&                               --& $ 1.00\pm0.11$ &  B, S, H \\
   9 &  J002015.04+591853.5 &  0.2 &$     2.00  \pm   0.36$&                               --& $ 0.88\pm0.11$ &  B, S    \\
  10 &  J002016.75+591556.2 &  0.8 &$     0.20  \pm   0.16$&                                            --& --&  H       \\
  11 &  J002020.99+591758.6 &  0.1 &$    12.40  \pm   0.82$&                  $ 0.79\pm0.05$ & $ 1.00\pm0.16$ &  H, B, S \\
  12 &  J002021.59+591901.5 &  0.5 &$     1.60  \pm   0.51$&                                            --& --&  B, H    \\
  13 &  J002022.92+591538.4 &  0.7 &$     0.27  \pm   0.17$&                                            --& --&  B       \\
  14 &  J002023.59+591444.0 &  0.1 &$     4.02  \pm   0.55$&                               --& $ 0.44\pm0.14$ &  B, S    \\
  15 &  J002024.54+591524.9 &  0.1 &$     7.41  \pm   0.66$&                  $ 0.13\pm0.10$ & $ 1.00\pm0.07$ &  B, H, S \\
  16 &  J002026.11+591844.7 &  0.5 &$     0.60  \pm   0.31$&                                            --& --&  S, B    \\
  17 &  J002026.68+591454.7 &  0.3 &$     0.54  \pm   0.22$&                                            --& --&  S, B    \\
  18 &  J002029.17+591651.2 &  0.0 &$   170.86  \pm   3.14$&                  $ 0.20\pm0.02$ & $ 0.94\pm0.01$ &  B, H, S \\
  19 &  J002030.74+591743.8 &  0.7 &$     0.31  \pm   0.17$&                                            --& --&  B, H    \\
  20 &  J002034.67+591556.8 &  0.2 &$     2.45  \pm   0.40$&                               --& $ 0.79\pm0.14$ &  B, S, H \\
  21 &  J002038.76+591740.7 &  0.3 &$     0.90  \pm   0.25$&                                            --& --&  B, H    \\
  22 &  J002040.14+591435.3 &  0.8 &$     0.67  \pm   0.46$&                                            --& --&  S       \\
  23 &  J002041.41+591712.4 &  0.9 &$     0.43  \pm   0.38$&                                            --& --&  H       \\
  24 &  J002046.45+591731.4 &  0.4 &$     0.61  \pm   0.48$&                                            --& --&  S, B    \\
  25 &  J002047.30+591320.2 &  0.7 &$     0.94  \pm   0.51$&                                            --& --&  B, S    \\
  26 &  J002050.06+591504.9 &  2.3 &$     0.37  \pm   0.46$&                                            --& --&  B       \\
  27 &  J002050.18+591408.7 &  0.8 &$     0.41  \pm   0.49$&                                            --& --&  B, H    \\
  28 &  J002052.81+591648.5 &  0.4 &$     1.11  \pm   0.52$&                                            --& --&  B, H, S \\
\enddata
\tablecomments{The definition of the bands:
0.3--0.7 (S1), 0.7--1.5 (S2), 1.5--3 (H1), and 3--7~keV (H2). 
In addition, S=S1+S2, H=H1+H2, and B=S+H.
 Column (1): Generic source number. (2): 
{\sl Chandra} X-ray Observatory (unregistered) source name, following the
{\sl Chandra} naming convention and the IAU Recommendation for Nomenclature
(e.g., http://cdsweb.u-strasbg.fr/iau-spec.html). (3): Position 
uncertainty (1$\sigma$) calculated from the maximum likelihood centroiding.  (4): On-axis source broad-band count rate --- the sum of the 
exposure-corrected count rates in the four
bands. (5-6): The hardness ratios defined as 
${\rm HR}=({\rm H-S2})/({\rm H+S2})$, and ${\rm HR1}=({\rm S2-S1})/{\rm S}$, 
listed only for values with uncertainties less than 0.2.
(7): The label ``B'', ``S'', or ``H'' mark the band in 
which a source is detected with the most accurate position that is adopted in
Column (3). 
}
  \end{deluxetable}
  \vfill
\begin{deluxetable}{lccrrrrr}
  \tabletypesize{\footnotesize}
  \tablecaption{Source Identification\label{tab_sou_id}}
  \tablewidth{0pt}
  \tablehead{
  \colhead{Source} &
  \colhead{2MASS ID Name} &
  \colhead{$\delta_{x,o}$ ($''$)} &
  \colhead{J, H, K$_s$ (mag)}
  }
  \startdata
   XP-1 & J001834.91+591947.6 &  1.2& 12.4 11.7 11.5 \\
   XP-3 & J001855.89+591240.6 &  0.8& 12.8 12.1 11.9 \\
   XP-9 & J001918.74+590820.3 &  0.9&  6.9  6.8  6.8 \\
  XP-11 & J001925.62+591419.1 &  1.7& 14.0 13.3 13.1 \\
  XP-12 & J001926.70+591105.4 &  2.1& 14.2 13.7 13.6 \\
  XP-16 & J001944.20+591812.5 &  3.6& 15.3 14.9 14.7 \\
  XP-19 & J001949.37+591333.7 &  2.6& 15.5 14.7 14.6 \\
  XP-21 & J001951.75+591327.1 &  1.5& 11.6 11.2 11.1 \\
  XP-34 & J002023.22+590814.0 &  1.2& 16.2 15.7 14.5 \\
  XP-35/XA-14 & J002023.51+591444.7 &  1.0&  8.8  8.5  8.5 \\
  XP-37/XA-16 & J002025.98+591845.3 &  1.3& 15.4 14.4 14.7 \\
 XP- 38 & J002026.57+590615.6 &  4.0& 15.1 14.6 14.3 \\
  XP-40 & J002034.39+591114.1 &  3.8& 15.1 14.6 14.3 \\
  XP-41 & J002033.80+591033.5 &  2.6& 13.0 12.2 12.0 \\
  XP-42/XA-20 & J002034.58+591557.3 &  3.9& 13.9 13.3 13.1 \\
  XP-43 & J002036.24+592405.6 &  3.0& 16.4 15.8 15.4 \\
  XP-45 & J002042.67+591851.4 &  1.1& 15.1 14.5 14.4 \\
  XP-47 & J002050.13+591525.9 &  2.6& 15.3 14.6 14.3 \\
  XP-50 & J002053.89+592108.2 &  2.0& 12.2 11.6 11.4 \\
  XP-52 & J002101.03+591104.5 &  2.2& 14.9 14.1 14.0 \\
  XP-53 & J002101.55+591518.1 &  1.8& 10.6  9.9  9.7 \\
  XP-56 & J002104.59+591539.9 &  0.9& 11.2 10.9 10.7 \\
  XP-58 & J002114.77+590845.5 &  1.6& 10.9 10.6 10.5 \\
 XP-60 & J002121.38+590905.3 &  0.6& 11.0 10.6 10.5 \\
  XP-61 & J002125.69+591904.9 &  2.1& 14.0 13.5 13.2 \\
  XP-65 & J002139.00+592243.3 &  0.3& 14.4 13.8 13.5 \\
  XP-71 & J002158.82+591327.2 &  0.7& 11.8 11.4 11.2 \\
  XP-73 & J002209.58+591401.1 &  3.8& 12.4 12.0 11.9 \\
XA-11 & J002020.90+591759.0 &  0.8& 16.4 15.4 15.5 \\
  XA-13 & J002022.80+591539.0 &  1.1& 13.5 13.2 13.1 \\
  XA-17 & J002026.61+591455.9 &  1.3& 13.7 13.0 12.7 \\
  XA-22 & J002040.10+591435.5 &  0.3& 10.8 10.4 10.2 \\
  XA-24 & J002046.44+591731.8 &  0.4& 15.4 14.8 14.4 \\
  XA-25 & J002047.26+591321.2 &  1.1& 14.8 14.1 13.9 \\
\enddata
  \end{deluxetable}
\end{document}